\documentclass[superscriptaddress,twocolumn, showkeys, showpacs]{revtex4}

\usepackage{amscd}
\usepackage{epsfig}
\usepackage{graphicx}
\usepackage{amsmath}
\usepackage{amssymb}
\usepackage{changebar}
\usepackage{longtable}

\newtheorem{definition}{Definition}

\newtheorem{theorem}{Theorem}
\newtheorem{corollary}{Corollary}

\newtheorem{conjecture}{Conjecture}

\newtheorem{condition}{Condition}
\newtheorem{sor}{Statement of Results}

\begin{document}
\bibliographystyle{apsrev}


\title{Structural Stability and Hyperbolicity Violation in High-Dimensional Dynamical Systems}
\author{D. J. Albers}
\email{albers@santafe.edu}
\affiliation{Physics Department, University of Wisconsin, Madison, WI 53706}
\affiliation{Santa Fe Institute, 1399 Hyde Park Road, Santa Fe, NM 87501}

\author{J. C. Sprott}
\email{sprott@physics.wisc.edu}
\affiliation{Physics Department, University of Wisconsin, Madison, WI 53706}

\date{\today}

\begin{abstract}
This report investigates the dynamical stability conjectures of Palis and Smale,
and Pugh and Shub from the standpoint of numerical observation and
lays the foundation for a stability conjecture.  As the dimension of a dissipative dynamical system is
increased, it is observed that the number of positive Lyapunov exponents increases
monotonically, the Lyapunov exponents tend towards continuous change
with respect to parameter variation, the number of observable periodic windows decreases (at least below
numerical precision), and a subset of parameter space exists such
that topological change is very common with small parameter
perturbation.  However, this seemingly inevitable topological
variation is never catastrophic (the dynamic type is preserved) if the
dimension of the system is high enough.     

\end{abstract}


\maketitle


                \def\thefootnote{\arabic{footnote}}
                        \setcounter{footnote}{0}







\section{Introduction}
\label{sec:introduction}
Much of the work in the fields of dynamical systems and differential equations have, for the last hundred years,
entailed the classification and understanding of the qualitative features of the
space of differentiable mappings.  A primary focus is
the classification of topological differences
between different systems (e.g. structural stability theory).  Of
course one of the primary difficulties is choosing a notion of
behavior that is not so strict that it differentiates on too
trivial a level, yet is strict enough that it has some meaning
(Palis-Smale used topological equivalence, Pugh-Shub use ergodicity).
The previous stability conjectures are with respect to any $C^r$ ($r
\geq 0$
varies from conjecture to conjecture)
perturbation allowing for variation of the mapping, both of the
functional form (with respect to the Whitney $C^r$ topology) and of parameter variation.  We will concern
ourselves with the latter issue.  Unlike much work involving stability
conjectures, our work is numerical, and it focuses on observable asymptotic
behaviors in high-dimensional systems.  Our chief claim is that generally, for
high-dimensional dynamical systems in our construction, there exist large
portions of parameter space such that topological variation inevitably
accompanies parameter variation, yet the topological variation
happens in a ``smooth,'' non-erratic manner.  Let us state our results
without rigor, noting that we will save more rigorous statements for
section (\ref{sec:conjectures}).
\label{sec:nnfunctionspace} \label{sec:dsfunctionspace}
\begin{sor}[Informal]
Given our particular impositions (sections (\ref{sec:ournnconstruction})
and (\ref{sec:dsembeddingconstruction})) upon the space of $C^r$ discrete-time maps
from compact sets to themselves, and an invariant measure (used for
calculating Lyapunov exponents), in the
limit of high dimension, there exists a subset of parameter space such
that strict hyperbolicity is violated on a nearly
dense (and hence unavoidable),
yet zero-measure (with respect to Lebesgue measure), subset of parameter space.
\end{sor}
A more refined version of this statement will contain all of our
results.  For mathematicians, we note that although the stability
conjecture of Palis and Smale \cite{globalanalysisss} is quite true
(as proved by Robbin \cite{robbinss},
Robinson \cite{robinsonss1}, and Ma\~n\'e \cite{maness}), we show that in high dimensions, this structural
stability may occur over such small sets in the parameter space that
it may never be observed in chaotic regimes of parameter space.
Nevertheless, this lack of observable structural stability has very mild
consequences for applied scientists.  

\subsection{Outline}
As this paper is attempting to reach a diverse readership, we will
briefly outline the work for ease of reading.  Of the remaining
introduction sections, section (\ref{sec:background}) can be skipped by readers
familiar with the stability conjecture of Smale and Palis and the
stable ergodicity of Pugh and Shub.

Following the introduction we will address various preliminary topics
pertaining to this report.  Beginning in section
(\ref{sec:dsembeddingconstruction}), we present the mathematical
justification for the study of time-delay maps being sufficient for a
general study of $d>1$ dimensional dynamical systems.  This section is
followed with a discussion of neural networks, beginning with their
definition in the abstract (section (\ref{sec:abstractnn})).  Following
the definition of neural networks, we explain the mappings neural networks are
able to approximate (section (\ref{sec:nnasfunctionapproximators})).  In
section (\ref{sec:ournnconstruction}) we give our specific construction
of neural networks.  Those uninterested in the mathematical justifications for our models
and only interested in our specific formulation should skip sections
(\ref{sec:dsembeddingconstruction}) thru (\ref{sec:nnasfunctionapproximators}) and concentrate on section
(\ref{sec:ournnconstruction}).  The discussion of our set of mappings is
followed by relevant definitions from hyperbolicity and ergodic theory
(section (\ref{sec:lceandhyperbolicity})).  It is here where we define
the Lyapunov spectrum, hyperbolic maps, and discuss relevant stability conjectures.  Section (\ref{sec:justificationtouselces}) provides
justification for our use of Lyapunov exponent calculations upon our
space of mappings (the neural networks).  Readers
familiar with topics in hyperbolicity and ergodic theory can skip this
section and refer to it as is needed for an understanding of the
results.  Lastly, in section (\ref{sec:definitionsfornumerics}), we make
a series of definitions we will need for our numerical arguments.
Without an understanding of these definitions, it is difficult to
understand both our conjectures of our arguments.

Section (\ref{sec:conjectures}) discusses the conjectures we wish to
investigate formally.  For those interested in just the
results of this report, reading sections
(\ref{sec:definitionsfornumerics}), (\ref{sec:conjectures}) and
(\ref{sec:summary}) will suffice.  The next section, section (\ref{sec:numericalerror}), discusses the
errors present in our chief numerical tool, the Lyapunov spectrum.
This section is necessary for a fine and careful understanding of this
report, but this section is easily skipped upon first reading.  We
then begin our preliminary numerical arguments.  Section
(\ref{sec:prelimnarguments}), addresses the three major properties we
need to argue for our conjectures.  For an understanding of our
arguments and why our conclusions make sense, reading this section
is necessary.  The main arguments regarding our
conjectures follow in section (\ref{sec:nargumentsofconjectures}).  It is in this section that we make the case for the claims of section
(\ref{sec:conjectures}).  The summary section (section (\ref{sec:summary})) begins with a summary
of our numerical arguments and how they apply to our conjectures.  We
then interpret our results in light of various stability conjectures
and other results from the dynamics community.

\subsection{Background}
\label{sec:background}
To present a full background with respect to the topics and
motivations for our study would be out of place in this report.  We
will instead discuss the roots of our problems and a few relevant
highlights, leaving the reader with references to the survey papers of
Burns et. al \cite{reviewstablergodicity}, Pugh and Shub
\cite{pughshubAMS}, Palis \cite{palisconjecture}, and Nitecki \cite{zbigniewbook} for a more thorough introduction.  

The origin of our work, as with all of dynamical systems,
lies with Poincar\'e who split the study of dynamics
in mathematics into two categories, conservative and dissipative
systems; we will be concerned with the latter.  We will refrain from beginning with Poincar\'e and
instead begin in the $1960's$ with the pursuit of the ``lost dream.''

The ``dream'' amounted to the conjecture that structurally stable
dynamical systems would be dense among all dynamical systems.  For mathematicians, the dream was
motivated primarily by a desire to classify dynamical
systems via their topological behavior.  For physicists and
other scientists however, this
dream was two-fold.  First, since dynamical systems (via differential
equations and discrete-time maps) are usually used to model physical
phenomena, a geometrical understanding of how these systems behave in
general is, from an intuitive standpoint, very insightful.  However, there is a more practical motivation for the stability
dream.  Most experimental scientists who work on highly nonlinear
systems (e.g. plasma physics and fluid dynamics) are painfully aware of
the existence of the dynamic stability that the mathematicians where
hoping to capture with the stability conjecture of Palis and  Smale.
When we write dynamic stability we do not mean fixed point versus
chaotic dynamics, rather we mean that upon normal
or induced experimental perturbations, dynamic types are highly
persistent.  Experimentalists have been attempting to control and
eliminate turbulence and chaos since they began performing
experiments --- it is clear from our experience that turbulence and chaos are highly
stable with respect to perturbations in highly complicated dynamical
systems, the why and how of the stability and what is the right notion
of equivalence to capture that stability is the question.  In a
practical sense, the hope lies in
that, if the geometric characteristics that allow chaos to
persist can be understood, it might be easier to control or even
eliminate those characteristics.  At the very least, it would be
useful to at least know very precisely why we can't control or rid our
systems of turbulent behavior.  At any rate, the dream was ``lost'' in the late $1960$'s via many
counter examples (\cite{globalanalysisss}), leaving room for a very
rich theory.  Conjectures regarding weaker forms of the dream for
which a subset of ``nice'' diffeomorphisms would be dense were put
forth, many lasted less than a day, and none worked.  The revival of
the dream in the $1990$'s involved a different notion of nice -
stable ergodicity.

Near the time of the demise of the "dream'' the notion of structural stability together with Smale's notion of
hyperbolicity was used to formulate the stability conjecture (the
connection between structural stability and hyperbolicity - now a theorem) \cite{palissmaless}.  The
stability conjecture says that ``a system is $C^r$ stable if its limit
set is hyperbolic and, moreover, stable and unstable manifolds meet
transversally at all points.'' \cite{palisconjecture} 

To attack the stability conjecture, Smale had introduced axiom A.
Dynamical systems that satisfy axiom A are strictly hyperbolic
(definition (\ref{definition:hyperbolicmap})) and
have dense periodic points on
the non-wandering set\footnote{$\Omega(f) = \{ x \in M | \forall
  \text{ neighborhood } U \text { of } x, \exists n \geq 0 \text{ such that }
f^n(U) \cap U \neq 0 \}$}.  A further condition that was needed is the
strong transversality condition - $f$ satisfies the strong
transversality condition when, on every $x \in M$, the stable and
unstable manifolds $W^s_x$ and $W^u_x$ are transverse at $x$.  That
axiom A and strong transversality imply
$C^r$ structural stability was shown by Robbin \cite{robbinss} for $r
\geq 2$ and Robinson \cite{robinsonss1} for $r=1$.  The other direction of the stability conjecture
was much more elusive, yet in $1980$ this was shown by 
Ma\~n\'e \cite{maness} for $r=1$.  

Nevertheless, due to many examples of structurally unstable systems
being dense amongst many ``common'' types of dynamical systems, proposing some global structure for a
space of dynamical systems became much more unlikely.  Newhouse \cite{newhousewild1} was able to show that infinitely many sinks occur
for a residual subset of an open set of $C^2$ diffeomorphisms near a
system exhibiting a homoclinic tangency.  Further, it was discovered
that orbits can be highly sensitive to initial conditions
\cite{lorenz}, \cite{milnorattractor}, \cite{sommererriddledbasins},
\cite{milnorkaneko}.  Much of the sensitivity to initial conditions was
investigated numerically by non-mathematicians.  Together, the
examples from both pure mathematics and the sciences sealed the demise
of the ``dream'' (via topological notions), yet they opened the door for
a wonderful and diverse theory.  Nevertheless, despite the fact that
structural stability does not capture all we wish it to capture, it is
still a very useful, intuitive tool.

Again, from a physical perspective, the question of the existence of
dynamic stability is not open - physicists and engineers have been
trying to suppress chaos and turbulence in high-dimensional systems
for several hundred years.  The trick in mathematics is writing
down a relevant notion of dynamic stability and then the relevant
necessary geometrical characteristics to guarantee dynamic stability.
From the perspective of modeling nature, structural stability says
that if one selects (fits) a model equation, small errors will be
irrelevant since small $C^r$ perturbations will yield topologically
equivalent models.  It is the topological equivalence that is too
strong a characteristic for structural stability to apply to the broad
range of systems we wish it to apply to.  Structural stability is
difficult to use in a
very practical way because it is very difficult to show (or disprove
the existence of) topological ($C^0$)
equivalence of a neighborhood of maps.  Hyperbolicity can be much
easier to handle numerically, yet it is not  always common.  Luckily,
to quote Pugh and Shub \cite{parthyppughshub}, ``a
little hyperbolicity goes a long way in guaranteeing stably ergodic
behavior.''  This thesis has driven the partial hyperbolicity branch
of dynamical systems and is our claim as well.  We will define
precisely what we mean by partial hyperbolicity and will discuss
relevant results \textit{a la} stable ergodicity and partial hyperbolicity.  

Our investigation
will, in a practical, computational context, investigate the extent to
which ergodic behavior and topological variation (versus parameter
variation) behave given a ``little bit'' of hyperbolicity.  Further,
we will investigate one of the overall haunting questions: how much of the space of
bounded $C^r$ $(r>0)$ systems is hyperbolic, and how many of the
pathologies found by Newhouse and others are observable (or even
existent) in the space of
bounded $C^r$ dynamical systems.  Stated more generally, how
does hyperbolicity (and thus structural stability) ``behave'' in a space of
bounded $C^r$ dynamical systems.


\section{Definitions and preliminaries}
\label{sec:preliminaries}
In this section we will define the following items: the family of
dynamical systems we wish to investigate; the function space we will
use in our experiments; Lyapunov exponents; and finally we will list definitions specific to our numerical
arguments.  The choice of scalar neural
networks as our maps of choice is motivated by their being ``universal
approximators.''

\subsection{Our space of mappings}
\label{sec:ourspaceofmappings}
The motivation and construction of the set of mappings we will use for our investigation of dynamical systems follows via two
directions, the embedding theorem of Takens (\cite{takensstit}, \cite{noa}) and the neural network
approximation theorems of Hornik, Stinchomebe, and White \cite{hor2}.
We will use the Takens embedding theorem to demonstrate how studying time-delayed maps of
the form $f:R^d \rightarrow R$ is a natural choice for studying
standard dynamical systems of the form $F: R^d \rightarrow R^d$.  This
is important as we will be using time-delayed scalar neural networks
for our study.  The neural network approximation theorems show that
neural networks of a particular form are open and dense in several
very general sets of functions and thus can be used to
approximate any function in the allowed function spaces.  

There is overlap, in a sense, between these two constructions.  The embedding
theory shows an equivalence or the approximation capabilities of scalar time-delay dynamics
with standard, $x_{t+1} = F(x_t)$ ($x_i \in R^d$) dynamics.  There is
no mention of, in a practical sense, the explicit functions in the
Takens construction.  The
neural network approximation results show in a precise and practical
way, what a neural network is, and what functions it can approximate.
It says that neural networks can approximate the
$C^r(R^d)$ mappings and their derivatives, but there is no mention  of
the time-delays we wish to use.  Thus we need to discuss both
the embedding theory and the neural network approximation theorems.   

Those not interested in the mathematical justification of our
construction may skip to section (\ref{sec:ournnconstruction}) where we
define, in a concrete manner, our neural networks.

\subsubsection{Dynamical systems construction}
\label{sec:dsembeddingconstruction}
We wish, in this report, to investigate dynamical systems on compact
sets.  Specifically, begin with a compact manifold $M$ of dimension $d$ and a
diffeomorphism $F \in C^r(M)$ for $r \geq 2$ defined as:
\begin{equation}
\label{equation:ds1}
x_{t+1} = F(x_t) 
\end{equation}
with $x_t \in M$.  However, for computational reasons, we will
be investigating this space with neural networks that can approximate
(see section (\ref{sec:nnasfunctionapproximators})) dynamical systems $f \in C^r(R^d, R)$ that are time-delay
maps given by:
\begin{equation}
\label{equation:ds2}
y_{t+1} = f(y_t, y_{t-1}, \dots, y_{t-(d-1)})
\end{equation}
where $y_t \in R$.  Both systems (\ref{equation:ds1}) and (\ref{equation:ds2}) form dynamical
systems.  However, since we intend to use systems of the
form (\ref{equation:ds2}) to investigate the space of dynamical
systems as given in equation (\ref{equation:ds1}), we must show
that a study of mappings of the form (\ref{equation:ds2}) is somehow
equivalent to mappings of the form (\ref{equation:ds1}).  We will
demonstrate this by employing an embedding
theorem of Takens to demonstrate the relationship between time-delay
maps and non-time-delay maps in a more general and formal setting.

We call $g \in C^k(M, R^n)$ an embedding if
$k \geq 1$ and if the local derivative map (the Jacobian - the first
order term in the Taylor expansion) is one-to-one for every point $x
\in M$ (i.e. $g$ must be an immersion).  The idea of the Takens
embedding theorem is that given a $d$-dimensional dynamical system and
a ``measurement function,'' $E: M \rightarrow R$ ($E$ is a $C^k$ map),
where $E$ represents some empirical style measurement of $F$, there is
a Takens map (which does the embedding) $g$ for which $x \in M$ can be
represented as a $2d+1$ tuple $(E(x), E \circ F(x), E \circ F^2(x), \dots, E
\circ F^{2d}(x))$ where $F$ is an ordinary difference equation
(time evolution operator) on $M$.  Note that the $2d+1$ tuple is a time-delay map
of $x$.  We can now state the Takens embedding theorem:  
\begin{theorem}{\bf (Takens' embedding theorem \cite{takensstit} \cite{noa})}
Let $M$ be a compact manifold with dimension $d$.  There is an open dense subset $S \subset \mathit{Diff}(M) \times
C^k(M, R)$ with the property that the Takens map
\begin{equation}
g:M \rightarrow R^{2d+1}
\end{equation}
given by $g(x) = (E(x), E \circ F(x), E \circ F^2(x), \dots, E
\circ F^{2d}(x))$ is an embedding of $C^k$ manifolds, when $(F,
E) \in S$.
\end{theorem} 
Here $\mathit{Diff}(M)$ is the space of $C^k$ diffeomorphisms from $M$ to
itself with the subspace topology from $C^k(M, M)$.  Thus, there is an equivalence between time-delayed Takens maps of
``measurements'' and the ``actual'' dynamical system operating in time
on $x_t \in M$.  This equivalence is that of an embedding (the
Takens map), $g:M \rightarrow R^{2d+1}$.

\begin{figure*}
\begin{center}
\epsfig{file=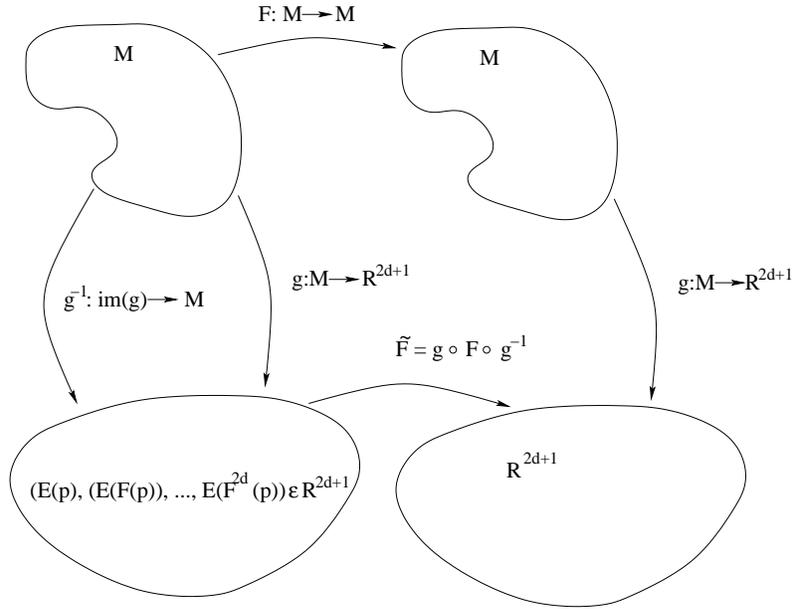, height=8cm}
\caption{Schematic diagram of the Takens embedding theorem and how it
  applies to our construction.}
\label{fig:takensfigure}
\end{center}
\end{figure*}

To demonstrate how this applies to our circumstances, consider figure
(\ref{fig:takensfigure}) in which $F$ and $E$ are as given above and the
embedding $g$ is explicitly given by:
\begin{equation}
g(x_t) = (E(x_t), E(F(x_t)), \dots, E(F^{2d}(x_t)))
\end{equation} 
In a colloquial, experimental sense, $\tilde{F}$ just keeps track of
the observations from the measurement function $E$, and, at each time
step, shifts the newest observation into the $2d+1$ tuple and
sequentially shifts the scalar observation at time $t$ ($y_t$) of the $2d+1$ tuple to the
$t-1$ position of the $2d+1$ tuple.  In more explicit notation, $\tilde{F}$ is the following mapping:
\begin{equation}
\label{equation:F-tilde}
(y_1, \dots, y_{2d+1}) \mapsto (y_2, \dots, y_{2d+1}, g(F(g^{-1}(y_1,
  \dots, y_{2d+1}))))
\end{equation}
where, again, $\tilde{F} = g \circ F \circ g^{-1}$.  The neural networks we will propose in the sections that follow can
approximate $\tilde{F}$ and its derivatives (to any order) to
arbitrary accuracy (a notion we will make more precise later).  

Let us summarize what we are attempting to do: we wish to investigate
dynamical systems given by (\ref{equation:ds1}), but for computational
reasons we wish to use dynamical systems given by  (\ref{equation:ds2}); the Takens embedding theorem says that dynamical systems of the form
(\ref{equation:ds1}) can be generically represented (via the Takens embedding map
$g$) by time-delay dynamical systems of the form (\ref{equation:F-tilde}).   Since neural networks will
approximate dynamical systems of the form (\ref{equation:F-tilde}) on a
compact and metrizable set, it will
suffice for our investigation of dynamical systems of the form
(\ref{equation:ds1}) to consider the space of neural networks
mapping compact sets to compact sets as given in section (\ref{sec:abstractnn}).

\subsubsection{Abstract neural networks}
\label{sec:abstractnn}
Begin by noting that, in general, a neural network is a $C^r$ mapping $\gamma: R^n
\rightarrow R$.  More specifically, the set of feedforward networks
with a single hidden layer, $\Sigma(G)$, can be written:
\begin{equation}
\label{equation:abstractnn}
\Sigma (G) \equiv \{ \gamma:R^d \rightarrow R | \gamma(x) = \sum_{i=1}^N \beta_i
G(\tilde{x}^T \omega_i) \}
\end{equation}
where $x \in R^d$, is the $d-$vector of networks inputs, $\tilde{x}^T
\equiv (1, x^T)$ (where $x^T$ is the transpose of $x$), $N$ is the
number of hidden units (neurons), $\beta_1, \dots, \beta_N \in R$ are
the hidden-to-output layer weights, $\omega_1, \dots, \omega_N \in
R^{d+1}$ are the input-to-hidden layer weights, and $G: R^d
\rightarrow R$ is the hidden layer activation function (or neuron).  The
partial derivatives of the network output function, $\gamma$, are 
\begin{equation}
\frac{\partial g(x)}{ \partial x_k} = \sum_{i=1}^N \beta_i \omega_{ik}
DG(\tilde{x}^T \omega_i)
\end{equation}
where $x_k$ is the $k^{th}$ component of the $x$ vector, $\omega_{ik}$
is the $k^{th}$ component of $\omega_i$, and $DG$ is the usual first
derivative of $G$.  The matrix of partial derivatives (the Jacobian) takes a particularly simple form when the $x$
vector is a sequence of time delays ($x_t = (y_t, y_{t-1}, \dots,
y_{t-(d-1)})$ for $x_t \in R^d$ and $y_i \in R$).  It is for precisely
this reason that we choose the time-delayed formulation.

\subsubsection{Neural networks as function approximations}
\label{sec:nnasfunctionapproximators}
We will begin with a brief description of spaces of maps useful
for our purposes and conclude with the keynote theorems of Hornik
et al. \cite{hor2} necessary for our work.  Hornik et al. provided the
theoretical justification for the use of neural networks as
function approximators.  The aforementioned authors provide a degree of generality that we will not
need; for their results in full generality see
\cite{hor}, \cite{hor2}.

The ability of neural networks to approximate functions which are of particular interest, can be most easily seen
via a brief discussion of Sobolev function space, $S_p^m$.  We
will be brief, noting references Adams \cite{adams} and Hebey \cite{hebey} for
readers wanting more depth with respect to Sobolev spaces.  For the sake of clarity and simplification, let us make a
few remarks which will pertain to the rest of this section:   
\begin{itemize}
\item[i.] $\mu$ is a measure; $\lambda$ is the standard Lebesgue
measure; for all practical purposes, $\mu = \lambda$;

\item[ii.] $l$, $m$ and $d$ are finite, non-negative integers; $m$ will be
  with reference to a degree of continuity of some function spaces, and $d$ will be the dimension of the space we are
  operating on;

\item[iii.] $p \in R$, $1 \leq  p < \infty$; $p$ will be with reference to a norm --- either the $L_p$ norm or
  the Sobolev norm; 

\item[iv.] $U \subset R^d$, $U$ is measurable.

\item[v.] $\alpha = (\alpha_1, \alpha_2, \dots, \alpha_d)^T$ is a
  $d$-tuple of non-negative integers (or a multi-index) satisfying $|\alpha| = \alpha_1 + \alpha_2 + \dots +
\alpha_k$, $|\alpha| \leq m$;

\item[vi.] for $x \in R^d$, $x^{\alpha} \equiv x_1^{\alpha_1} \cdot x_2^{\alpha_2}
\dots x_d^{\alpha_d}$.

\item[vii.] $D^{\alpha}$ denotes the partial derivative of order $|\alpha|$
\begin{equation}
\frac{\partial^{|\alpha|}}{\partial x^{\alpha}} \equiv
\frac{\partial^{|\alpha|}}{(\partial x_1^{\alpha_1} \partial
  x_2^{\alpha_2} \dots \partial x_d^{\alpha_d} )}
\end{equation}

\item[viii.] $u \in L_{loc}^1(U)$ is a locally integrable, real valued function on
$U$

\item[ix.] $\rho^m_{p, \mu}$ is a metric, dependent on the subset
  $U$, the measure $\mu$, and $p$ and $m$ in a manner we will define
  shortly;

\item[x.] $\parallel \cdot \parallel_p$ is the standard norm in $L_p(U)$;

\end{itemize}

Letting $m$ be a positive integer and $1 \leq  p < \infty$, we define the
Sobolev norm, $\parallel \cdot \parallel_{m, p}$, as follows:
\begin{equation}
||u||_{m, p} = \left( \sum_{0 \leq |\alpha| \leq m} (
  \parallel D^{\alpha}u \parallel_p^p) \right)^{1/p}
\end{equation}
where $u \in L_{loc}^1(U)$ is a locally integrable, real valued function on
$U \subset R^d$ ($u$ could be significantly more general) and $|| \cdot
||_p$ is the standard norm in $L_p(U)$.  Likewise, the Sobolev metric
can be defined:
\begin{equation}
\rho_{p, \mu}^m(f, g) \equiv || f - g ||_{m, p, U, \mu}
\end{equation}
It is important to note that this metric is dependent on $U$.

For ease of notation, let us define the set of $m$-times
differentiable functions on $U$,
\begin{equation}
C^m(U) = \{ f \in C(U) | D^{\alpha} f \in C(U), ||D^{\alpha} f||_p < \infty \forall \alpha, |\alpha| \leq m \}
\end{equation}
We are now free to define the Sobolev space for which our results will
apply.
\begin{definition}
For any positive integer $m$ and $1 \leq p < \infty$, we define a
Sobolev space $S^m_p(U, \lambda)$ as the vector space on which $|| \cdot
||_{m,p}$ is a norm:
\begin{equation}
S_p^m(U, \lambda) = \{ f \in C^m(U) | \text{  } ||D^{\alpha}f||_{p, U, \lambda}
< \infty \text{ for all } |\alpha| \leq m \}
\end{equation}
Equipped with the Sobolev norm, $S_p^m$ is a Sobolev space over $U
\subset R^d$.
\end{definition}

Two functions in $S_p^m(U, \lambda)$ are close in the Sobolev metric
if all the derivatives of order $0 \leq |\alpha| < m$ are close in the
$L_p$ metric.  It is useful to recall that we are attempting to approximate $\tilde{F} = g \circ F \circ
g^{-1}$ where $\tilde{F}: R^{2d+1} \rightarrow R$; for this task the functions from
$S_p^m(U, \lambda)$ will serve us quite nicely.  The whole point of all this machinery is to state approximation theorems that
require specific notions of density.  Otherwise we would refrain and
instead use the standard notion of $C^k$ functions --- the functions that
are $k$-times differentiable uninhibited by a notion of a metric or norm.

Armed with a specific function space for
which the approximation results apply (there are many more), we will conclude this section by
briefly stating one of the approximation results.  However, before stating the
approximation theorem, we need two definitions --- one which
makes the notion of closeness of derivatives more precise and one
which gives the sufficient conditions for the activation functions to
perform the approximations.

\begin{definition}{ \bf ($m$-uniformly dense) }
Assume $m$ and $l$ are non-negative integers $0 \leq m \leq l$, $U
\subset R^d$, and $S \subset C^l(U)$.  If, for any $f \in S$, compact
$K \subset U$, and $\epsilon>0$ there exists a $g \in \Sigma(G)$ such
that:
\begin{equation}
\max_{|\alpha| \leq m} \sup_{x \in K} |D^{\alpha}f(x) -
D^{\alpha}g(x)| < \epsilon
\end{equation}
then $\Sigma(G)$ is $m$-uniformly dense on compacta in $S$.
\end{definition}
It is this notion of $m$-uniformly dense in $S$ that provides all the
approximation power of both the mappings and the derivatives (up to
order $l$) of the mappings.  Next we will supply the condition on our activation function
necessary for the approximation results.
\begin{definition}{ \bf ($l$-finite)}
Let $l$ be an non-negative integer.  $G$ is said to be $l$-finite for $G
\in C^l(R)$ if:
\begin{equation}
\label{equation:lfinite}
0< \int |D^lG| d \lambda < \infty
\end{equation}
i.e. the $l^{th}$ derivative of $G$ must be both bounded away from
zero, and finite for all $l$ (recall $d \lambda$ is the standard Lebesgue
volume element).
\end{definition}
The hyperbolic tangent, our activation function, is $l$-finite.

With these two notions, we can state one of the many existing
approximation results.
\begin{corollary}{ \bf (corollary $3.5$ \cite{hor2} )}
If $G$ is $l$-finite, $0 \leq m \leq l$, and $U$ is an open subset of
$R^d$, then $\Sigma(G)$ is $m$-uniformly dense on compacta in
$S_p^m(U, \lambda)$ for $1 \leq p < \infty$.
\end{corollary}
In general, we wish to investigate differentiable mappings of compact
sets to themselves.  Further, we wish for the derivatives to be finite
almost everywhere.  Thus the space $S_p^m(U, \lambda)$ will suffice for
our purposes.  Our results also apply to piecewise differentiable
mappings.  However, this requires a more general Sobolev space, $W^m_p$.
We have refrained from delving into the definition of this space since it
requires a bit more formalism, for those interested see \cite{hor2}
and \cite{adams}.

\subsubsection{Our neural network construction}
\label{sec:ournnconstruction}
The single layer feed-forward neural networks ($\gamma$'s from the above
section) we will consider are of the form
\begin{equation}
\label{equation:net1}
 x_{t} = \beta_0 + \sum_{i=1}^{N}{{\beta}_i G \left( s{\omega}_{i0} + s
 \sum_{j=1}^{d}{{\omega}_{ij} x_{t-j} } \right)} 
\end{equation}
which is a map from $R^{d}$ to $R$.  The squashing function
$G$, for our purpose, will be the hyperbolic tangent.  In (\ref{equation:net1}), $N$ represents the number of hidden units or
neurons, $d$ is the input or embedding dimension of the system which
functions simply as the number of time lags, and $s$ is a scaling factor on the
weights.

The parameters are real (${\beta}_i , {w}_{ij} , x_{j} , s  \in {R}$) and the $\beta_{i}$'s and ${w}_{ij}$'s are elements of weight matrices (which we
hold fixed for each case).  The initial conditions are denoted as $({x}_{0}, x_{1}, \ldots , x_{d})$, and $({x}_{t}, x_{t+1}, \ldots
, x_{t+d})$ represent the current state of
the system at time $t$.  

We assume that the $\beta$'s are $iid$ uniform over $[0,1]$
and then re-scaled to satisfy the condition
$\sum_{i=1}^{N}{{{\beta}_{i}^{2}}} = N$.  The ${w}_{ij}$'s are $iid$ normal with zero mean and unit variance.
The $s$ parameter is a real number and can be interpreted as the
standard deviation of the $w$ matrix of weights.  The initial $x_j$'s
are chosen $iid$ uniform on the interval $[-1,1]$.  All the weights and
initial conditions are selected randomly using a pseudo-random number
generator \cite{lec}, \cite{pres}.

We would like to make a few notes with respect to our squashing function, $\tanh()$.  First, $\tanh(x)$, for
$|x| \gg 1$ will tend to behave much like a binary
function.  Thus, the states of the neural network will tend toward the
finite set $( \beta_0 \pm \beta_1 \pm \beta_2 \dots \pm \beta_N)$, or a
set of $2^N$ different states.  In the limit where the arguments of
$\tanh()$ become infinite, the neural network will have periodic
dynamics.  Thus, if $<\beta>$ or $s$ become very large, the system
will have a greatly reduced dynamic variability.  Based on
this problem, one might feel tempted to bound the $\beta$'s a la
$\sum_{i=1}^{N}{|{\beta}_i |} = k$ fixing $k$ for all $N$ and $d$.
This is a bad idea however since, if the $\beta_{i}$'s are restricted to a
sphere of radius $k$, as $N$ is increased, $\langle
{\beta_{i}}^2 \rangle$
goes to zero \cite{albers1}.  The other extreme of
our squashing also yields a very specific behavior type.  For $x$ very
near $0$, the $\tanh(x)$ function is nearly linear.  Thus choosing $s$ small will force the
dynamics to be mostly linear, again yielding fixed point and periodic
behavior (no chaos).  Thus the scaling parameter $s$ will provide a
unique bifurcation parameter that will sweep from linear ranges to
highly non-linear ranges, to binary ranges - fixed points to chaos and
back to periodic phenomena.

Note that in a very practical sense, the measure we are imposing
on the set of neural networks is our means of selecting the
weights that define the networks.  This will
introduce a bias into our results that is unavoidable in such
experiments; the very act of picking networks out of the space will
determine, to some extent, our results.  Unlike actual physical experiments, we could, in principle,
prove an invariance of our results to our induced measure.  This is
difficult and beyond the scope of this paper.  Instead it will suffice
for our purposes to note specifically what our measure is (our weight
selection method), and how it
might bias our results.  Our selection method will include all
possible networks, but clearly not with the same likelihood.  In the absence of a theorem with
respect to an invariance of our induced measure, we must be careful
in stating what our results imply about the ambient function space. 

\subsection{Characteristic Lyapunov exponents and Hyperbolicity}
\label{sec:lceandhyperbolicity}
Let us now define the
diagnostics for our numerical simulations.  We will begin by defining
structural stability and its relevant notion of topological equivalence
(between orbits, attractors, etc), topological conjugacy.  We will then discuss
notions that are more amenable to a numerical study, yet can be
related to the geometrical notions of structural stability.
Hyperbolicity will be defined in
three successive definitions, each with increasing generality, culminating with
a definition of partial hyperbolicity.  This will be followed with a
global generalization of local eigenvalues, the Lyapunov spectrum.  We will include here a brief statement regarding the
connection between structural stability, hyperbolicity, and the Lyapunov spectrum.

\begin{definition}{ \bf (Structural Stability)}
A $C^r$ discrete-time map, $f$, is structurally stable if there is a $C^r$
neighborhood, $V$ of $f$, such that any $g \in V$ is topologically conjugate
to $f$, i.e. for every $g \in V$, there exists a homeomorphism $h$ such that $f = h^{-1}
\circ g \circ h$.
\end{definition}
In other words, a map is structurally stable if, for all other maps
$g$ in a $C^r$ neighborhood, there exists a homeomorphism that will
map the domain of $f$ to the domain of $g$, the range of $f$ to the
range of $g$, and the inverses respectively.  This is a purely
topological notion.    

Next, let us begin by defining hyperbolicity in an intuitive manner,
followed by a more general definition useful for our purposes.  Let us start
with a linear case: 
\begin{definition}{ \bf (Hyperbolic linear map)}
\label{defintion:hyperbolicfixedpoint}
A linear map of $R^n$ is called hyperbolic if all of its eigenvalues
have modulus different from one.
\end{definition}

The above definition can be generalized as follows:
\begin{definition}{ \bf( Hyperbolic map)}
\label{definition:hyperbolicmap}
A discrete-time map $f$ is said to be hyperbolic on a compact
invariant set $\Lambda$ if there exists a continuous splitting of the
tangent bundle, $TM|_{\Lambda} = E^s \oplus E^u$, and there are
constants $C>0$, $0<\lambda <1$, such that $||Df^n|_{E_x^s}||< C
\lambda^n$ and $||Df^{-n}|_{E_x^u}||< C \lambda^n$ for any $n>0$ and
$x\in \Lambda$.
\end{definition}
Here the stable bundle $E^s$ (respectively unstable bundle $E^u$) of $x
\in \Lambda$ is the set of points $p \in M$ such that $|f^k(x) -
f^k(p)| \rightarrow 0$ as $k \rightarrow \infty$ ($k \rightarrow -
  \infty$ respectively).

As previously mentioned, strict hyperbolicity is a bit
restrictive; thus let us make precise the notion of a ``little bit''
of hyperbolicity:
\begin{definition}{ \bf (Partial hyperbolicity)}
\label{definition:partialhyperbolicity}
The diffeomorphism $f$ of a smooth Riemannian manifold $M$ is said to
be partially hyperbolic if for all $x \in M$ the
tangent bundle $T_xM$ has the invariant splitting:
\begin{equation}
T_xM = E^u(x) \oplus E^c(x) \oplus E^s(x)
\end{equation}
into strong stable $E^s(x) = E^s_f(x)$, strong unstable $E^u(x) =
E^u_f(x)$, and central $E^c(x) = E^c_f(x)$ bundles, at least two of
which are non-trivial\footnote{If $E^c$ is trivial, $f$ is simply
  Anosov, or strictly hyperbolic.}.  Thus there will exist numbers
$0 < a < b < 1 < c < d$ such that, for all $x \in M$:  
\begin{eqnarray}
v \in E^u(x) \Rightarrow d||v|| \leq ||D_xf(v)|| \\
v \in E^c(x) \Rightarrow b||v|| \leq ||D_xf(v) || \leq c ||v|| \\
v \in E^s(x) \Rightarrow ||D_xf(v)|| \leq a ||v||
\end{eqnarray}
\end{definition}
%
%
More specific characteristics and definitions can be found
in references \cite{brinpesinparthyp}, \cite{hirshpughshub}, \cite{parthyppughshub},
\cite{reviewstablergodicity}, and \cite{partialhyplese}.  The key
provided by definition \ref{definition:partialhyperbolicity} is the
allowance of center bundles, zero Lyapunov exponents, and in general,
neutral directions, which are not allowed in strict
hyperbolicity.  Thus we are allowed to keep the general framework of
good topological structure, but we lose structural stability.  With
non-trivial partial hyperbolicity (i.e. $E^c$ is not null), stable
ergodicity replaces structural
stability as the notion of dynamic stability in the Pugh-Shub
stability conjecture (conjecture
(\ref{conjecture:pughshubstabilityconjecture}) of \cite{parthyppughshub2}).  Thus
what is left is to again attempt to show the extent to which stable
ergodicity persists, and topological variation is not pathological,
under parameter variation with non-trivial center bundles present.
Again, we note that results in this area will be discussed in a later section.

In numerical simulations we
will never observe an orbit on the unstable, stable, or center
manifolds.  Thus we will need a
global notion of stability averaged along a given orbit (which will exist under
weak ergodic assumptions).  The notion we seek is captured by the
spectrum of Lyapunov exponents.

We will initially define Lyapunov
exponents formally, followed by a more practical, computational
definition. 
\begin{definition}{ \bf (Lyapunov Exponents)}
\label{definition:lyapunovexponentsdef}
Let $f:M \rightarrow M$ be a diffeomorphism (i.e. discrete time map)
on a compact Riemannian manifold of dimension $m$.  Let $| \cdot |$ be
the norm on the tangent vectors induced by the Riemannian metric on
$M$.  For every $x \in M$ and $v \in T_x M$ Lyapunov exponent at $x$
is denoted:
\begin{equation}
\label{equation:lces}
\chi(x, v) = \lim \sup_{t \rightarrow \infty} \frac{1}{t} log ||Df^n v||
\end{equation}
\end{definition}
Assume the function $\chi(x, \cdot)$ has only finitely many values on
$T_x M \ \{ 0 \}$ (this assumption may not be true for our dynamical
systems) which we denote $\chi_1^f(x) < \chi_2^f(x) \dots <
\chi_m^f(x)$.  Next denote the filtration of $T_x M$
associated with $\chi(x, \cdot)$, $\{ 0 \} = V_0(x) \subsetneqq V_1(x)
\subsetneqq \dots \subsetneqq V_m (x) = T_x M$, where $V_i(x) = \{ v
\in T_x M | \chi(x, v) \leq \chi_i(x) \}$.  The number $k_i = dim(V_i(x))
- dim(V_{i-1}(x))$ is the multiplicity of the exponent $\chi_i(x)$.
In general, for our networks over the parameter range we are
considering, $k_i = 1$ for all $0 < i \leq m$.  Given the above, the
Lyapunov spectrum for $f$ at $x$ is defined:
\begin{equation}
\label{eqn:lcespectrum}
\text{Sp} \chi(x) = \{ \chi_j^k(x) | 1 \leq i \leq m \}
\end{equation}
(For more information regarding Lyapunov exponents and spectra see
\cite{pesinlebook}, \cite{katokbook}, or \cite{manedsbook}.

A more computationally
motivated formula for the Lyapunov exponents is given as:
\begin{equation}
\label{equation:practicallce}
\chi_j = \lim_{N \rightarrow \infty} \frac{1}{N} \sum_{k=1}^N
\ln(\langle (Df_k \cdot \delta x_j)^T ,(Df_k \cdot \delta x_j) \rangle )
\end{equation}
where $\langle , \rangle$ is the standard inner product, $\delta x_j$
is the $j^{th}$ component of the $x$ variation\footnote{In a practical
sense, the $x$ variation is the initial separation or perturbation of $x$.}
and $Df_k$ is the ``orthogonalized'' Jacobian of $f$ at the $k^{th}$
iterate of $f(x)$.  Through the course of our discussions we will
dissect equation (\ref{equation:practicallce}) further.  It should
also be noted that Lyapunov exponents have been shown to be
independent of coordinate system, thus the specifics of our above
definition do not affect the outcome of the exponents.

The existence of Lyapunov exponents is established by a
multiplicative ergodic theorem (for a nice example, see theorem
($1.6$) in \cite{ruelleetheory}).  There exist many such theorems for
various circumstances.  The first multiplicative ergodic theorem was
proven by Oseledec \cite{oseledec}; many others - \cite{katok80},
\cite{ruellehilbert}, \cite{ruelleetheory}, \cite{pesinmet1}, \cite{pesinmet2}, \cite{pesinmet3}, and \cite{brinpesinparthyp} -  have
subsequently generalized his original result.  We will refrain from
stating a specific multiplicative ergodic theorem; the conditions
necessary for the existence of Lyapunov exponents are exactly the
conditions we place on our function space in section
in (\ref{sec:justificationtouselces}).  In other words, a $C^r$ ($r>0$) map of a compact
manifold $M$ to itself and an $f-$invariant probability measure
$\rho$, on $M$.  For specific treatments we leave the curious reader
to study the aforementioned references, noting that our construction
follows from \cite{ruellehilbert}, \cite{brinpesinparthyp}, and \cite{partialhyplese}.

There is an intimate relationship between Lyapunov exponents and
global stable and unstable manifolds.  In fact, each Lyapunov exponent
corresponds to a global manifold.  We will be using the global
manifold structure as our measure of topological equivalence, and the
Lyapunov exponents to classify this global structure.  Positive
Lyapunov exponents correspond to global unstable manifolds, and negative
Lyapunov exponents correspond to global stable manifolds.  We will
again refrain from stating the existence theorems for these global
manifolds, and instead note that in addition to the requirements for
the existence of Lyapunov exponents, the existence of global
stable/unstable manifolds corresponding the negative/positive Lyapunov
exponents requires $Df$ to be injective.  For
specific global unstable/stable manifold theorems see \cite{ruellehilbert}.

The theories of hyperbolicity, Lyapunov exponents and structural stability have had a
long, wonderful, and tangled history  (for good starting points see
\cite{smaledds} or \cite{anosov1}).  We will, of course, not scratch
the surface with our current discussion, but rather put forth the
connections relevant for our work.  Lyapunov exponents are the logarithmic
average of the (properly normalized) eigenvalues of the local
(linearization at a point) Jacobian along a
given orbit.  Thus for periodic orbits, the Lyapunov exponents are
simply the log of the eigenvalues.  A periodic orbit with period $p$ is
hyperbolic if either the eigenvalues of the time $p$ map are not one, or the Lyapunov
exponents are not zero.  The connection between structural stability
and hyperbolocity is quite beautiful and has a long and wonderful
history beginning with Palis and Smale \cite{smalessnotdense}.  For purposes of
interpretation later, it will be useful to state the solution of the stability conjecture:

\begin{theorem}{\bf (Ma\~n\'e \cite{maness} theorem $A$, Robbin
    \cite{robbinss}, Robinson \cite{robinsonss})}
A $C^1$ diffeomorphism (on a compact, boundaryless
manifold) is structurally stable if and only if it satisfies axiom A and the strong transversality
condition.
\end{theorem}

Recall that axiom A says the diffeomorphism is hyperbolic with
dense periodic points on its
non-wandering set $\Omega$ ($p \in \Omega$ is non-wandering if for any neighborhood $U$ of $x$, there is
an $n>0$ such that $f^n(U) \cap U \neq 0$).  We will save a further explicit discussion of this interrelationship for
a later section, noting that much of this report investigates the
above notions and how they apply to our set of maps.

Finally, for a nice, sophisticated introduction to the above topics
see \cite{katokbook}.

\subsection{Conditions needed for the existence of Lyapunov exponents}
\label{sec:justificationtouselces}
Lyapunov exponents are one of our principal diagnostics, thus we must
briefly justify their existence for our construction.  We will begin
with a standard construction for the existence and
computation of Lyapunov exponents as defined by the theories of Katok \cite{katok80},
Ruelle \cite{ruellehilbert}, \cite{ruelleetheory}, Pesin
\cite{pesinmet1}, \cite{pesinmet2}, \cite{pesinmet3}, Brin and
Pesin \cite{brinpesinparthyp}, and  Burns, Dolgopyat and Pesin \cite{partialhyplese}.  We will then note how this applies to
our construction.  (For more practical approaches to the numerical
calculation of Lyapunov spectra see \cite{benn1}, \cite{benn2},
\cite{shimnag}, and \cite{deelepaper}.)

Let $\mathcal{H}$ be a separable real Hilbert space (for practical purposes
$R^n$), and let $X$ be an open subset of $\mathcal{H}$.  Next let $(X, \Sigma,
\rho)$ be a probability space where $\Sigma$ is a $\sigma-$algebra of
sets, and $\rho$ is a probability measure, $\rho(X) = 1$ (see
\cite{loeve} for more information).  Now consider a $C^r$ ($r>1$)
map $f_t : X \mapsto X$ which preserves $\rho$ ($\rho$ is
$f-$invariant) defined for $t \geq T_0 \geq 0$ such
that $f_{t_1 + t_2} = f_{t_{1}} \circ f_{t_2}$ and that $(x, t)
\mapsto f_t (x)$, $Df_t(x)$ is continuous from $X \times [T_0, \infty)$
to $X$ and bounded on $\mathcal{H}$.  Assume that $f$ has a compact invariant
set 
\begin{equation}
\label{equation:invariantset}
\Lambda = \{ \bigcap_{t>T_0} f_t (X) | f_t(\Lambda) \subseteq \Lambda \}
\end{equation}
and $Df_t$ is a compact bounded operator for $x \in \Lambda$, $t>T_0$.  Finally, endow $f_t$
with a scalar parameter $s \in [0: \infty]$.  This gives us
the space (a metric space - the metric will be defined heuristically
in section \ref{sec:ournnconstruction}) of one parameter, $C^r$ measure-preserving maps from bounded compact sets to
themselves with bounded first derivatives.  It is for a space of the
above mappings that Ruelle shows the existence of Lyapunov exponents
\cite{ruellehilbert} (similar, requirements are made by Brin and
Pesin \cite{brinpesinparthyp} in a slightly more general setting).     

Now we must quickly justify our use of Lyapunov exponents.  Clearly, we can take $X$ in the above construction to be the $R^d$ of
section (\ref{sec:dsembeddingconstruction}).  As our neural networks map their domains to
compact sets, and they are constructed as time-delays, their domains
are also compact.  Further, their derivatives are bounded up to
arbitrary order, although for our purposes, only the first order need be
bounded.   Because the neural networks are deterministic and bounded,
there will exist an invariant set of some type.  All we need yet deal
with is the measure preservation of which previously there is no
mention.  This issue is partially addressed in
\cite{dee_reconstruction} in our neural network context.  There
remains much work to achieve a full
understanding of Lyapunov exponents for general dissipative dynamical
systems that are not absolutely continuous, for a current treatment
see \cite{pesinlebook}.  The specific measure theoretic properties of
our networks (i.e. issues such as absolute continuity,
uniform/non-uniform hyperbolicity, basin structures, etc) is a topic
of current investigation.

\subsection{Definitions for numerical arguments}
\label{sec:definitionsfornumerics}
Since we are conducting a numerical experiment, we will present some notions
needed to test our conjectures numerically.  We will begin with a notion of continuity.  The heart of continuity is
based on the following idea: if a neighborhood about a point in the
domain is shrunk, this implies a shrinking of a neighborhood of the
range.  However, we do not have infinitesimals at
our disposal.  Thus, our statements of numerical continuity
will necessarily have a statement regarding the limits of numerical
resolution below which our results are uncertain. 

Let us now begin with a definition of bounds on the domain and range:
\begin{definition}{ \bf( $\epsilon_{num}$)}
$\epsilon_{num}$ is the numerical accuracy of a Lyapunov exponent, $\chi_j$.
\end{definition}

\begin{definition}{ \bf ($\delta_{num}$)}
$\delta_{num}$ is the numerical accuracy of a given parameter under variation.  
\end{definition}

Now, with our $\epsilon_{num}$ and $\delta_{num}$ defined as our numerical limits
in precision, let us define numerical continuity of Lyapunov
exponents.    
\begin{definition}{ \bf ($num-$continuous Lyapunov exponents)}
\label{definition:ncontinuity}
Given a one parameter map $f: R^1 \times R^d \rightarrow R^d$, $f \in C^r$, $r>0$, for
which characteristic exponents $\chi_j$ exist (and are the same under all
invariant measures).  The map $f$ is said to have $num$-continuous
Lyapunov exponents at $(\mu, x) \in R^1 \times R^d$ if for
$\epsilon_{num}>0$ there exists a $\delta_{num}>0$  such that if:
\begin{equation}
|s - s'| < \delta_{num}
\end{equation}
then
\begin{equation}
|\chi_j(s) - \chi_j(s')| < \epsilon_{num}
\end{equation}   
for $s, s' \in R^1$, for all $j \in N$ such that $0 < j \leq d$.  
\end{definition}
Another useful definition related to continuity is that of a function
being Lipschitz continuous.
\begin{definition}{ \bf ($num-$Lipschitz)}
\label{definition:nlipschitz}
Given a one parameter map $f: R^1 \times R^d \rightarrow R^d$, $f \in C^r$, $r>0$, for
which characteristic exponents $\chi_j$ exist (and are the same under all
invariant measures), the map $f$ is said to have $num$-Lipschitz
Lyapunov exponents at $(\mu, x) \in R^1 \times R^d$ if there exists a
real constant $0 < k_{\chi_j}$ such that
\begin{equation}
|\chi_j(s) - \chi_j(s')| < k_{\chi_j} |s-s'|
\end{equation} 
Further, if the constant $k_{\chi_j}<1$, the Lyapunov exponent is said
to be contracting\footnote{Note, there is an important difference
  between the Lyapunov exponent contracting, which implies some sort
  of convergence to a particular value, versus a negative Lyapunov
  exponent that implies a contracting direction on the manifold or in
  phase space.} on the interval $[s, s']$ for all $s'$ such that $|s-s'|<\delta_{num}$.
\end{definition}
Note that neither of these definitions imply strict continuity, but rather,
they provide bounds on the difference between
the change in parameter and the change in Lyapunov exponents.  It is important to note that these notions are highly
localized with respect to the domain in consideration.  We will not
imply some sort of global continuity using the above definitions,
rather, we will
use these notions to imply that Lyapunov exponents will continuously (within
numerical resolution) cross
through zero upon parameter variation.  We can
never numerically prove that Lyapunov exponents don't jump across
zero, but for most computational exercises, a jump across zero that is
below numerical precision is not relevant.  This notion of continuity
will aid in arguments regarding the existence of periodic windows in
parameter space.   

Let us next define a Lyapunov exponent zero-crossing:
\begin{definition}{ \bf (Lyapunov exponent zero-crossing)}
A Lyapunov exponent zero-crossing is simply the point $s_{\chi_j}$ in parameter
space such that a Lyapunov exponent continuously (or
$num-$continuously) crosses zero.  e.g. for $s-\delta$, $\chi_i > 0$,
and for $s + \delta$, $\chi_i <0$.  
\end{definition}

For this report, a Lyapunov exponent zero-crossing is a transverse
intersection with the real line.  For our networks  non-transversal
intersections of the Lyapunov exponents with the real line certainly
occur, but for the portion of parameter space we are investigating,
they are extremely rare.  Along the route-to-chaos for our networks,
such non-transversal intersections are common, but will save the
discussion of that topic for a different report.  Orbits for which the Lyapunov spectrum can be defined (in a
numerical sense, Lyapunov exponents are defined when they are
convergent), yet at least one of the exponents is zero are
called non-trivially $num-$partially hyperbolic.  We must be careful making
statements with respect to the existence zero Lyapunov exponents
implying the existence of corresponding center manifolds $E^c$ as we
can do with the positive and negative exponents and their respective
stable and unstable manifolds.


Lastly, we define a notion of denseness for a numerical context.
There are several ways of achieving such a notion --- we will use the
notion of a dense sequence.  
\begin{definition}{ \bf ($\epsilon$-dense)} 
Given an $\epsilon>0$, an open interval $(a, b) \subset R$, and a sequence $\{ c_1, \dots, c_n
\}$, $\{ c_1, \dots, c_n
\}$ is $\epsilon$-dense in $(a, b)$ if there exists an $n$ such that for any $x \in (a, b)$, there
is an $i$, $1 \leq i < n$, such that $\text{dist}(x, c_i) < \epsilon$.
\end{definition}  

In reality however, we will be interested in a
sequence of sequences that are ``increasingly'' $\epsilon$-dense in an
interval $(a, b)$.  In other words, for the sequence of sequences
$$\begin{array}{ccc}
  c_1^1, & \cdots, & c_{n_1}^1 \\
  c_1^2, & \cdots, & c_{n_2}^2 \\
  \vdots & \vdots & \vdots \\
\end{array} $$
where $n_{i+1} > n_i$ (i.e. for a sequence of sequences with
increasing cardinality), the subsequent sequences for increasing $n_i$ become a closer approximation of an $\epsilon$-dense
sequence.  Formally ---
\begin{definition}{ \bf (Asymptotically Dense ($a-$dense))}
\label{definition:eventuallydense}
A sequence $S_j = \{ c_1^j, \dots, c_{n_j}^j \} \subset (a, b)$ of
finite subsets is asymptotically dense in $(a, b)$, if for any
$\epsilon > 0$, there is an $N$ such that $S_j$ is $\epsilon$-dense
if $j\geq N$.
\end{definition}
For a intuitive example of this, consider a sequence $S$ of $k$
numbers where $q_k \in S$, $q_k \in (0, 1)$.  Now
increase the cardinality of the set, spreading elements in such a way
that they are uniformly distributed over
the interval.  Density is achieved with the cardinality of infinity,
but clearly, with a finite but arbitrarily high number of elements, we
can achieve any approximation to a dense set that we wish.
There are, of course, many ways we can have a countably infinite set that is not dense, and, as we are working with numerics, we must concern ourselves
with how we will approach this asymptotic density.  We now need a
clear understanding of when this definition will apply to a given set.  There are many
pitfalls; for instance, we wish to avoid sequences such as $(1,
\frac{1}{2}, \frac{1}{3}, \cdots, \frac{1}{n}, \cdots)$.  We will, in
the section that addresses $a-$density, 
state the necessary conditions for an $a-$dense set for our purposes.

\section{Conjectures}
\label{sec:conjectures}
The point of this exercise is verifying three properties of
$C^r$ maps along a one-dimensional interval in parameter space.  The
first property is the existence of a
collection of points along an interval in parameter space such that
hyperbolicity of the mapping is violated.  The second property, which is really dependent upon the first and
third properties, is the existence of an interval in parameter space of
positive measure such that topological change (in the sense of changing numbers of unstable
manifolds) with respect to slight parameter variation on the
aforementioned interval is common.  The final property we wish to show, which will be crucial for arguing
the second property, is that on the aforementioned interval in
parameter space, the topological change will not yield periodic windows
in the interval if the dimension of the mapping is sufficiently
high.  More specifically, we will show that the ratio of periodic
window size to parameter variation size ($\delta_s$) goes to zero on our chosen interval. 

\begin{condition}
\label{condition:ass2}
Given a map (neural network) as defined in section
(\ref{sec:ournnconstruction}), if the parameter $s \in R^1$ is varied
$num-$continuously, then the Lyapunov exponents vary $num-$continuously.
\end{condition}

There are many counterexamples to this condition, so many
of our results will rest upon our ability to show how generally the
above condition applies in high-dimensional systems.

\begin{definition}{ \bf (Chain link set)}
\label{definition:chainlinkset}
Assume $f$ is a mapping (neural network) as defined in section
(\ref{sec:ournnconstruction}).  A \textbf{chain link set} is denoted:
\begin{align*}
V = \{ s \in R \ | \ & \text{$\chi_j(s) \neq 0$ for all $0 < j  \leq d$} \\
  & \text{ and $\chi_j(s) > 0$ for some $j>0$} \}
\end{align*}
\end{definition}
If $\chi_j(s)$ is continuous at its Lyapunov exponent zero-crossing, as we will show later (\textit{a la} condition (\ref{condition:ass2})), then $V$
is open.  Next, let $C_k$ be a connected component of the closure
of $V$, $\overline{V}$.  It can be shown that $C_k \cap \ V$ is a union of
disjoint, adjacent open intervals of the form
$\bigcup_i (a_i, a_{i+1})$.  
\begin{definition}{ \bf (Bifurcation link set)}
Assume $f$ is a mapping (neural network) as defined in section
(\ref{sec:ournnconstruction}).  Denote a \textbf{bifurcation link set} of $C_k \cap V$ as:
\begin{equation}
V_i = (a_i, a_{i+1})
\end{equation}
\end{definition}
Assume the number of positive Lyapunov exponents for each $V_i \subset
V$ remains constant, if, upon a monotonically increasing variation in the parameter $s$, the number of positive Lyapunov for
$V_i$ is greater than the number of positive Lyapunov exponents for
$V_{i+1}$, $V$ is said to be LCE decreasing.  Specifically, the
endpoints of $V_i$'s are the points where there exist Lyapunov exponent zero crossings.  We are
not particularly interested in these sets however, rather we are interested in the
collection of endpoints adjoining these sets.
\begin{definition}{ \bf (Bifurcation chain subset)}
\label{definition:bifurcationchainsubset}
Let $V$ be a chain link set, and $C_k$ a connected component of
$\overline{V}$.  A \textbf{bifurcation chain subset} of $C_k \cap V$ is denoted:
\begin{equation}
U_k = \{ a_i \}
\end{equation}
or equivalently:
\begin{equation}
U_k = \partial(C_k \cap V)
\end{equation}
\end{definition}

\begin{figure*}
\begin{center}
\epsfig{file=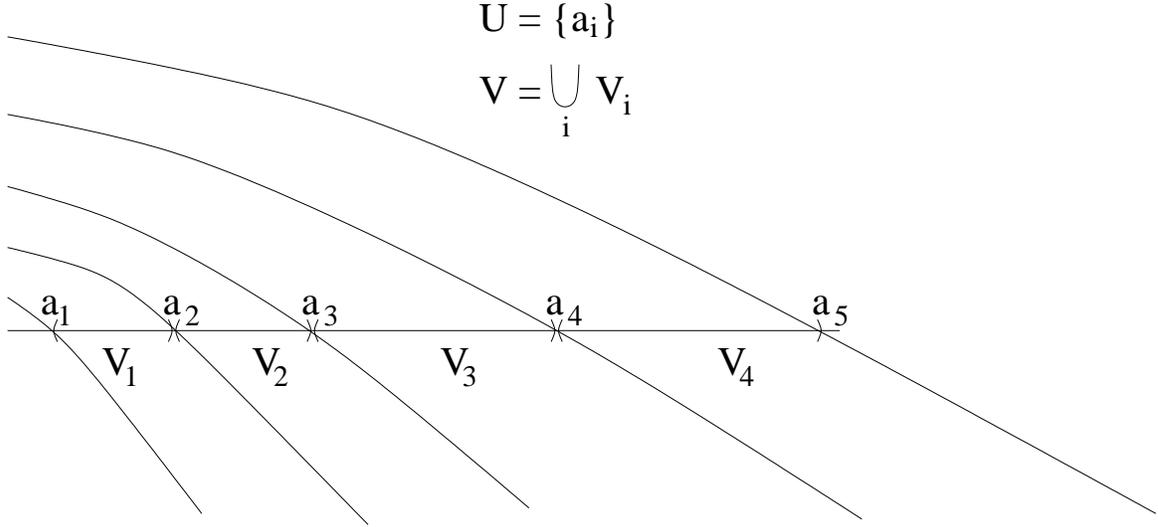, height=7cm}
\caption{An intuitive diagram for chain link sets, $V$, bifurcation link
  sets, $V_i$, and bifurcation chain sets, $U$. for an LCE decreasing chain link
  set $V$. }
\label{fig:chainsets}
\end{center}
\end{figure*}

For our purposes in this work, we will consider a bifurcation chain
subset $U$ such that $a_1$ corresponds to the last zero
crossing of the least positive exponent and $b_n$ will depend upon the
specific case and dimension.  In a practical sense, $a_1 \sim 0.5$ and
$b_n \sim 6$.  For higher dimensional networks, $b_n \sim 6$ will
correspond to a much higher $n$ than for a low-dimensional network.
For an intuitive picture of what we wish to depict with the above
definitions, consider figure (\ref{fig:chainsets}).

We will now state the conjectures, followed by some definitions and an
outline of what we will test and why those tests will verify our
claims.

\begin{conjecture}{ \bf (Hyperbolicity violation)}
\label{conjecture:hypvol}
Assume $f$ is a mapping (neural network) as defined in section
(\ref{sec:ournnconstruction}) with a sufficiently high number of
dimensions, $d$.  There exists at least one bifurcation chain subset $U$. 
\end{conjecture}

The intuition arises from a straightforward consideration of
the neural network construction in section
(\ref{sec:ournnconstruction}).  From consideration of our specific neural networks and their
activation function, $\tanh()$, it is clear that variation of the scaling
parameter, $s$, on the variance of the interaction weights \textbf{$\omega$} forces the neural
networks from a linear region, through a non-linear region, and into a
binary region.  This implies that, given a neural network that is
chaotic for some value of $s$, upon the monotonically increasing variation of $s$ 
from zero, the dynamical behavior will begin at a fixed point,
proceed through a sequence of bifurcations, become chaotic, and
eventually become periodic.  If the number of positive Lyapunov exponents can be shown to
increase with the dimension of the network and if the Lyapunov exponents can be shown to
vary relatively continuously with respect to parameter variation with
increasing dimension, then there will be many points along the
parameterized curve such that there will exist neutral directions.  The ideas listed above provide the framework for computational
verification of conjecture (\ref{conjecture:zerobifvol}).  We must
investigate conjecture (\ref{conjecture:hypvol}) with respect to
the subset $U$ becoming $a-dense$ in its closure and the existence of
very few (ideally a single) connected components of $\overline{V}$.

\begin{conjecture}{ \bf (Existence of a Codimension $\epsilon$ bifurcation set)}
\label{conjecture:zerobifvol}
Assume $f$ is a mapping (neural network) as defined in section
(\ref{sec:ournnconstruction}) with a sufficiently high number of
dimensions, $d$, and a bifurcation chain set $U$ as per conjecture
(\ref{conjecture:hypvol}).  The two following (equivalent) statements hold:  
\begin{itemize}
\item[i.] In the infinite-dimensional limit, the cardinality of $U$ will go to
infinity, and the length $\max |a_{i+1} - a_i|$ for all $i$ will tend
to zero on a one dimensional interval in parameter space.  In other
words, the bifurcation chain set $U$ will be $a-$dense in its
closure, $\overline{U}$.

\item[ii.] In the
asymptotic limit of high dimension, for all $s \in U$, and for
all $f$ at $s$, an arbitrarily small perturbation $\delta_s$ of $s$ will produce
a topological change.  The topological change will correspond to a different number of global
stable and unstable manifolds for $f$ at $s$ compared to $f$ at $s +
\delta$.
\end{itemize}

\end{conjecture}

Assume $M$ is a $C^r$ manifold of topological dimension $d$ and $N$ is a
submanifold of $M$.  The codimension of $N$ in $M$ is defined $\mathit{codim}(N) =
dim(M) - dim(N)$.  If there exists a curve $p$ through $M$ such that
$p$ is transverse to $N$ and the $\mathit{codim}(N) \leq 1$, then there
will not exist an arbitrarily small perturbation to $p$ such that $p$
will become non-transverse to $N$.  Moreover, if
$\mathit{codim}(N)=0$ and $p \bigcap N \subset \mathit{int}(N)$, then
there does not even exist an arbitrarily small perturbation of $p$ such that
$p$ intersects $N$ at a single point of $N$, i.e. the intersection
cannot be made non-transverse with an arbitrarily small perturbation. 

\begin{figure}
\begin{center}
\epsfig{file=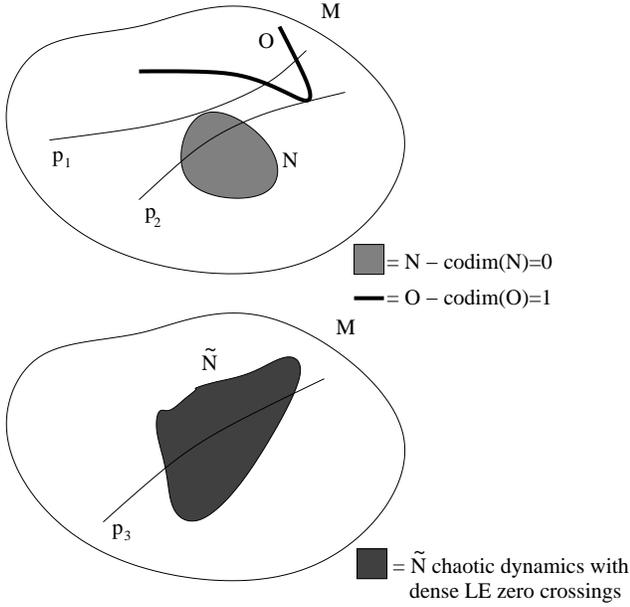, height=8cm}
\caption{The top drawing represents various standard pictures from
  transversality theory.  The bottom drawing represents an idealized
  version (in higher dimensions) of transversality catering to our
  arguments. }
\label{fig:codimension}
\end{center}
\end{figure}

The former paragraph can be more easily understood via figure
(\ref{fig:codimension}) where we have drawn four different
circumstances.  This first circumstance, the curve $p_1 \cap N$, is an
example of a non-transversal intersection with a codimension $0$ submanifold.
This intersection can be perturbed away with an arbitrarily small
perturbation of $p_1$.  The intersection, $p_2 \cap N$, is a
transversal intersection with a codimension $0$ submanifold, and this
intersection cannot be perturbed away with an arbitrarily small
perturbation of $p_2$.  Likewise, the intersection, $p_1 \cap O$,
which is an example of a transversal intersection with a codimension $1$
submanifold cannot be made non-transverse or null via an arbitrarily small
perturbation of $p_1$.  The intersection $p_2 \cap O$ is a
non-transversal intersection with a codimension $1$ submanifold and can be
perturbed away with an arbitrarily small perturbation of $p_2$.  This
outlines the avoid-ability of codimension $0$ and $1$ submanifolds
with respect to curves through the ambient manifold $M$.  The point is
that non-null, transversal intersections of curves with codimension
$0$ or $1$ submanifolds cannot be made non-transversal with
arbitrarily small perturbations of the curve.  Transversal intersections of curves with codimension $2$
submanifolds, however, can always be removed by an arbitrarily small
perturbation due to the existence of a ``free'' dimension.  A
practical example of such would be the intersection of a curve with
another curve in $R^3$ --- one can always pull apart the two curves
simply by ``lifting'' them apart.

In the circumstance proposed in conjecture
(\ref{conjecture:zerobifvol}), the set $U$ ($\tilde{N}$ in the
Fig. (\ref{fig:codimension})) will always have codimension $d$ because $U$ consists of
finitely many points, thus  any intersection with $U$ can be
removed by an arbitrarily small perturbation.  The point is that, as
$U$ becomes $a$-dense in $\bar{U}$,  $p_3 \bigcap \bar{U} = 0$
becomes more and more unlikely and the perturbations required to
remove the intersections of $p_3$ with $U$ (again, $\tilde{N}$ as in the
Fig. (\ref{fig:codimension}) ) will become more and more bizarre.  For a low-dimensional example, think of a ball of radius $r$ in $R^3$ that is populated by a
finite set of evenly distributed points, denoted $S_i$, where $i$ is
the number of elements in $S_i$.  Next fit a curve $p$ through
that ball in such a way that $p$ does not hit any points in
$S_i$.  Now, as the cardinality of $S_i$ becomes large, if $S_i$ is
$a$-dense in the ball of radius $r$, for the intersection of $p$ with
$S_i$ to remain null, the $p$ will need to become increasingly kinky.
Moreover, continuous, linear transformations of $p$ will become
increasingly unlikely to preserve $p \cap S_i = 0$.  It is
this type  of behavior  with respect to parameter variation that we
are arguing for with conjecture (\ref{conjecture:zerobifvol}).  However, figure
(\ref{fig:codimension}) is should only be used as an tool for
intuition --- our conjectures are with respect to a
particular interval in parameter space and not a general curve in
parameter space, let alone a family of curves or a high-dimensional
surface.  Conjecture (\ref{conjecture:zerobifvol}) is a first step
towards a more complete argument with respect to the above scenario.  For more information for where the above picture originates,
see \cite{soyomayorbifsets} or \cite{globalwiggins}.

To understand roughly why
we believe conjecture (\ref{conjecture:zerobifvol}) is reasonable, first take condition (\ref{condition:ass2}) for granted (we will expend
some effort showing where condition (\ref{condition:ass2}) is reasonable).  Next assume
there are arbitrarily many Lyapunov exponents near $0$ along some
interval of parameter space and that the Lyapunov exponent
zero-crossings can be shown to be $a-$dense with increasing dimension.
Further, assume that on the aforementioned interval, $V$ is LCE
decreasing.  Since varying the parameters continuously on some
small interval will move Lyapunov
exponents continuously, small changes in the parameters will guarantee
a continual change in the number of positive Lyapunov exponents.  One might think of this
intuitively relative to the parameter space as the set of Lyapunov
exponent zero-crossings forming a codimension $0$ submanifold with
respect to the particular interval of parameter space.  However, we will never
achieve such a situation in a rigorous way.  Rather, we will have an $a-$dense
bifurcation chain set $U$, which will have codimension $1$ in $R$ with
respect to topological dimension.  As the dimension of $f$ is increased, $U$ will behave more like a
codimension $0$ submanifold of $R$.  Hence the
metaphoric language, codimension $\epsilon$ bifurcation set.  The set $U$
will always be a codimension one submanifold as it is a finite set of
points.  Nevertheless, if $U$ tends toward being dense in its closure, it
will behave increasingly like a codimension zero submanifold.  This argument will not work for the entirety of the parameter
space, and thus we will show where, to what extent, and under what
conditions $U$ exists and how it behaves as the dimension of the
network is increased.

\begin{conjecture}{ \bf (Periodic window probability decreasing)}
\label{conjecture:windowsize}
Assume $f$ is a mapping (neural network) as defined in section
(\ref{sec:ournnconstruction}) and a bifurcation chain set $U$ as per conjecture
(\ref{conjecture:hypvol}).  In the
asymptotic limit of high dimension, the length of the bifurcation
chain sets, $l = |a_n - a_1|$, increases such that the cardinality of $U
\rightarrow m$ where $m$ is the maximum number of positive Lyapunov
exponents for $f$.  In other words, there will exist an interval in
parameter space (e.g. $s \in (a_1, a_n) \sim (0.1,4)$) where the probability of the existence
of a periodic window will go to zero (with respect to Lebesgue measure
on the interval) as the dimension becomes large.
\end{conjecture}

This conjecture is somewhat difficult to test for a specific function
since adding inputs completely changes the function.  Thus the curve
through the function space is an abstraction we are not afforded by
our construction.  We will save a more complete analysis (e.g. a search
for periodic windows along a high-dimensional surface in parameter
space) of conjecture
(\ref{conjecture:windowsize}) for a different report.  In this work,
conjecture (\ref{conjecture:windowsize}) addresses a very practical
matter, for it implies the existence of a much smaller number of
bifurcation chain sets.  The previous conjectures allow for the
existence of many of these bifurcation chains sets, $U$, separated by
windows of periodicity in parameter space.  However, if these windows
of periodic dynamics in parameter space vanish, we could end up with
only one bifurcation chain set --- the ideal situation for our
arguments.  We will not claim such, however we will
claim that the length of the set $U$ we are concerning
ourselves with in a practical sense will increase with increasing
dimension, largely due to the disappearance of periodic windows on
the closure of $V$.  With respect to this report, all that needs be shown is
that the window sizes along the path in parameter space for a
variety of neural networks decreases with increasing dimension.  From a
qualitative analysis it will be somewhat clear that the above
conjecture is reasonable.  

If this were actually making statements we could rigorously prove, conjectures (\ref{conjecture:hypvol}),
(\ref{conjecture:zerobifvol}), and (\ref{conjecture:windowsize}) would
function as lemmas for conjecture (\ref{conjecture:nongenercityss}).
\begin{conjecture}
\label{conjecture:nongenercityss} 
Assume $f$ is a mapping (neural network) as defined in section
(\ref{sec:ournnconstruction}) with a sufficiently high number of
dimensions, $d$, a bifurcation chain set $U$ as per conjecture
(\ref{conjecture:hypvol}), and the chain link set $V$.  The
perturbation size $\delta_s$ of $s \in C_{max}$, where $C_{max}$ is the largest
connected component of $\overline{V}$, for which $f|_{C_k}$ remains
structurally stable goes to zero as $d \rightarrow \infty$.
\end{conjecture}

Specific cases and the lack of density of structural
stability in certain sets of dynamical systems has been proven long
ago.  These examples were, however, very specialized and carefully
constructed circumstances and do not speak to the
commonality of structural stability failure.  Along the road
to investigating conjecture (\ref{conjecture:nongenercityss}) we will show that structural
stability will not, in a practical sense, be observable for a large set of very
high-dimensional dynamical systems along certain, important intervals in parameter
space even though structural stability is a property
that will exist on that interval with probability one (with respect to
Lebesgue measure).  To some, this conjecture might appear to contradict some well-known results in stability theory.  A
careful analysis of this conjecture, and its relation to known
results will be discussed in sections (\ref{sec:genssconclusion}) and (\ref{sec:usandss}).

The larger question that remains, however, is whether conjecture
(\ref{conjecture:nongenercityss}) is valid on high-dimensional surfaces
in parameter space.  We believe this is a much more difficult question
with a much more complicated answer.  We can, however, speak to a highly
related problem, the problem of whether chaos persists in high-dimensional dynamical systems.  Thus, let us now make a very imprecise conjecture that we will make more
concise in a later section.
\begin{conjecture}
\label{conjecture:robustchaos}
Chaos is a robust, high-probability behavior for high-dimensional,
bounded, nonlinear dynamical systems.
\end{conjecture}
This is not a revelation (as previously mentioned, many experimentalists
have been attempting to break this robust, chaotic behavior for the
last hundred years), nor is it a particularly precise statement.  We have studied this question using neural networks
much like those described in section (\ref{sec:ournnconstruction}), and we found that
for high-dimensional networks with a sufficient degree of
nonlinearity, the probability of chaos was near unity \cite{dechertjcandme}.  Over
the course of investigation of the above claims, we will see a
qualitative verification of conjecture (\ref{conjecture:robustchaos}).
A more complete study will come from combining results from this study
with a statistical perturbation study and combined with a study of
windows proposed by \cite{yorkenilpotency} and the closing lemma of
Pugh \cite{closinglemma}.

\section{Numerical errors in Lyapunov exponent calculation}
\label{sec:numericalerror}
Before we commence with our numerical arguments for the above
conjectures, we analyze the numerical errors
for both insight into how our chief diagnostic works and to establish bounds of
accuracy on the numerical results that will follow.  We will proceed
first with an analysis of single networks of varying dimensions,
providing intuition into the evolution of the calculation of the
Lyapunov spectrum versus iteration time.  We will follow this
analysis with a statistical study of $1000$ networks, measuring the
deviation from the mean of the exponent over $10000$ time steps, thus
noting how the individual exponents converge and to what extent the
exponents of all the networks converge.

\begin{figure*}
\begin{center}
\epsfig{file=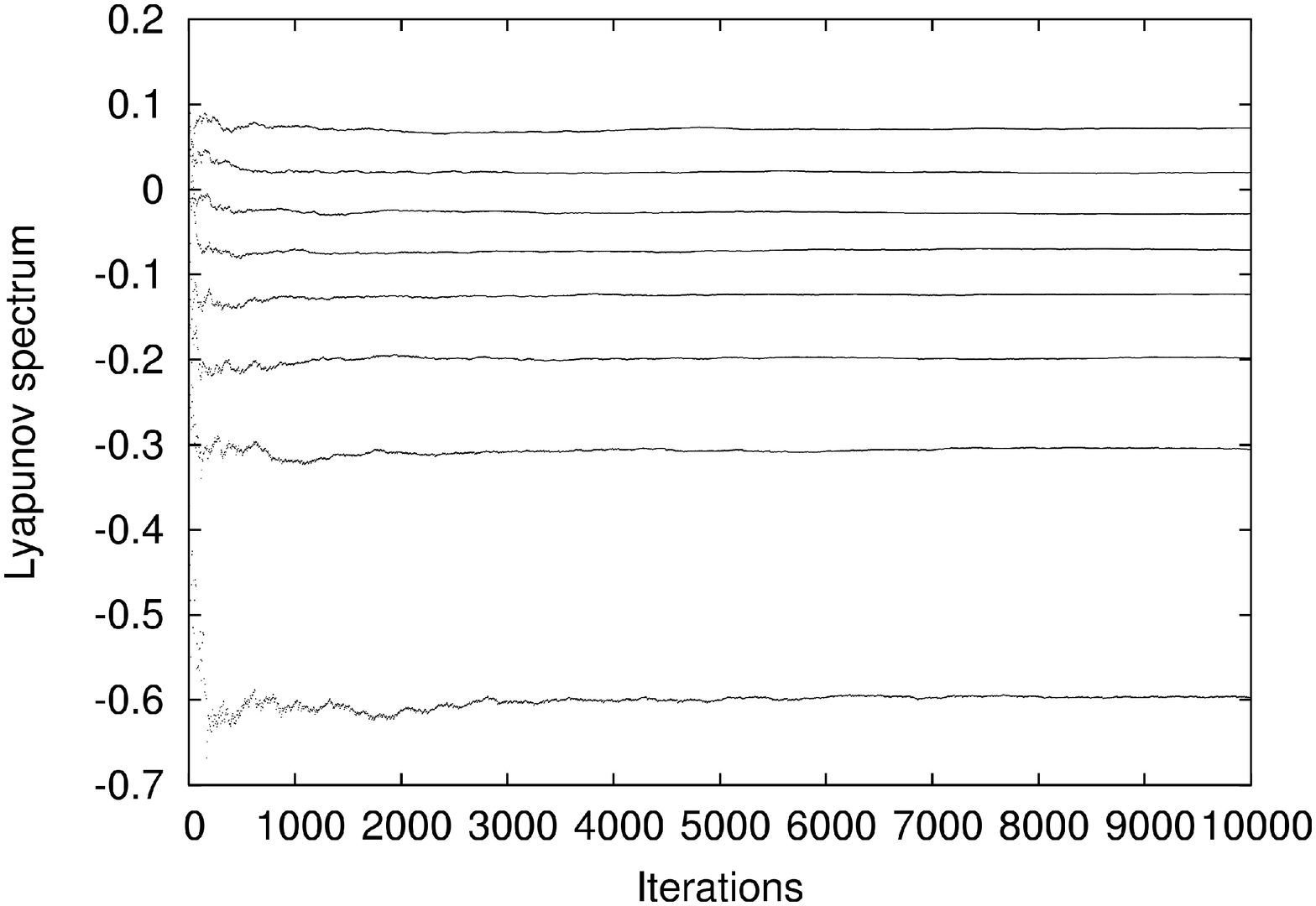, height=2.9in, angle=270}
\epsfig{file=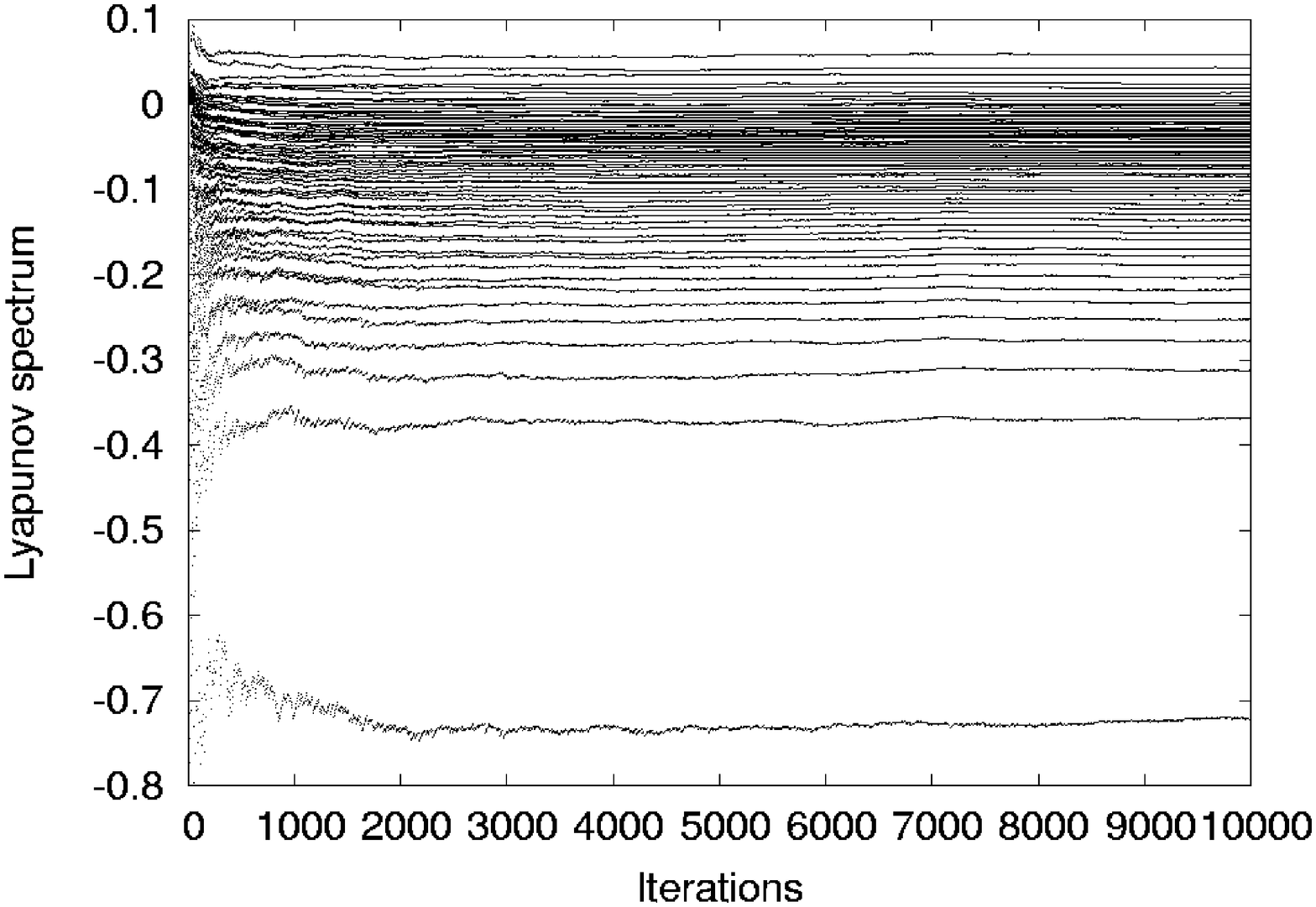, height=2.9in, angle=270}
\caption{LE spectrum versus iteration for individual networks with $32$
  neurons and $16$ (left, only the largest $8$ are shown) and $64$ (right) dimensions}
\label{fig:lcevsiteration16and64}
\end{center}
\end{figure*}

\begin{figure}
\begin{center}
\epsfig{file=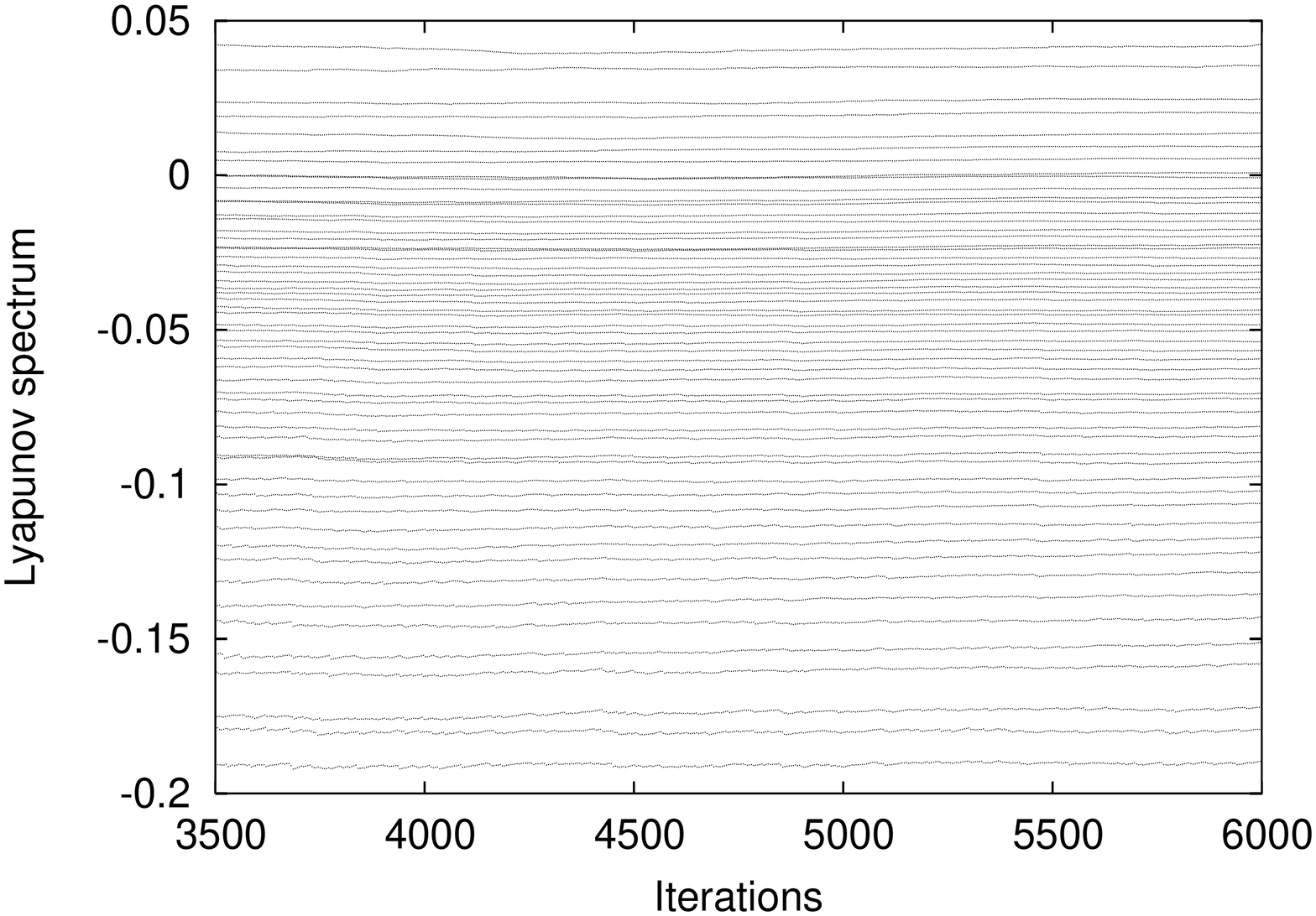, height=7cm, angle=270}
\caption{Close-up of LE spectrum versus iteration: 32 neurons, 64 dimensions}
\label{fig:closeup64}
\end{center}
\end{figure}

We will begin by considering Fig. (\ref{fig:lcevsiteration16and64}),
plots of the Lyapunov spectrum versus the first $10000$ iterations for
two networks with $16$ and $64$
dimensions.  After approximately $3000$ time steps, all
the large transients have essentially vanished, and aside from slight variation (especially on a time
scale long compared with a single time-step) the exponents appear to
have converged.  For the case with $16$
dimensions the exponents also appear to have converged.  The resolution for the
network with $64$ dimensions is not fine enough to verify a
distinction between exponents, thus consideration of Fig.
(\ref{fig:closeup64}) demonstrates clearly that the exponents converge
well within the inherent errors in the calculation, and are entirely
distinct for time steps
greater than $5500$ time steps.  It is worth noting that there are
times when very long term transients occur in our networks.  These
transients would not be detectable from the figures we have presented,
but these problem cases usually only exist near bifurcation points.
For the cases we are considering, these convergence issues do not seem
to affect our results\footnote{When an exponent is very nearly zero it
can tend to fluctuate above and below zero, but it is always very near
zero.  Thus although it might be difficult to resolve zero exactly ---
which is to be expected --- the
exponent is clearly very near zero which is all that really matters
for our  purposes.}.          

\begin{figure*}
\begin{center}
\epsfig{file=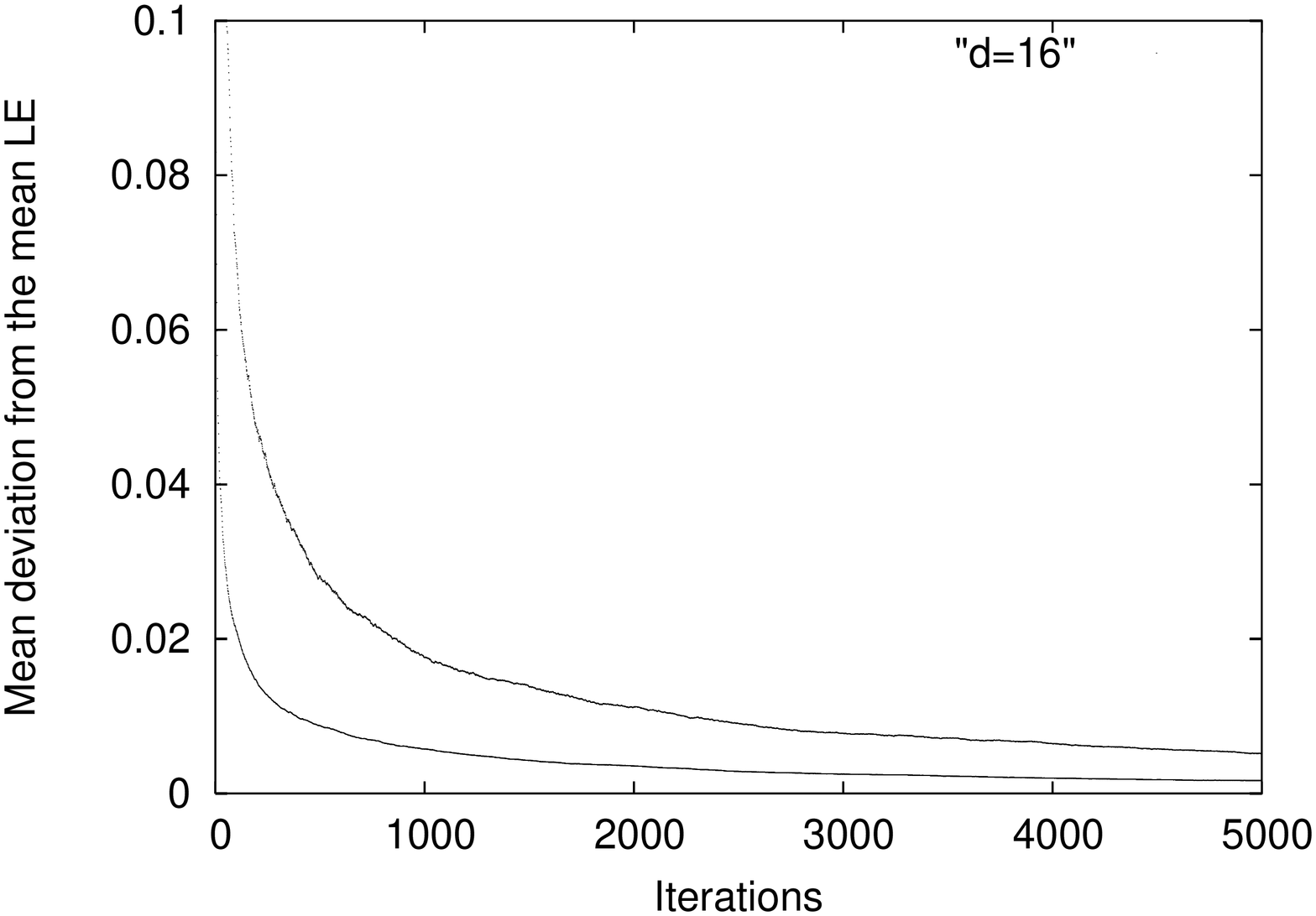, height=2.3in, angle=270}
\epsfig{file=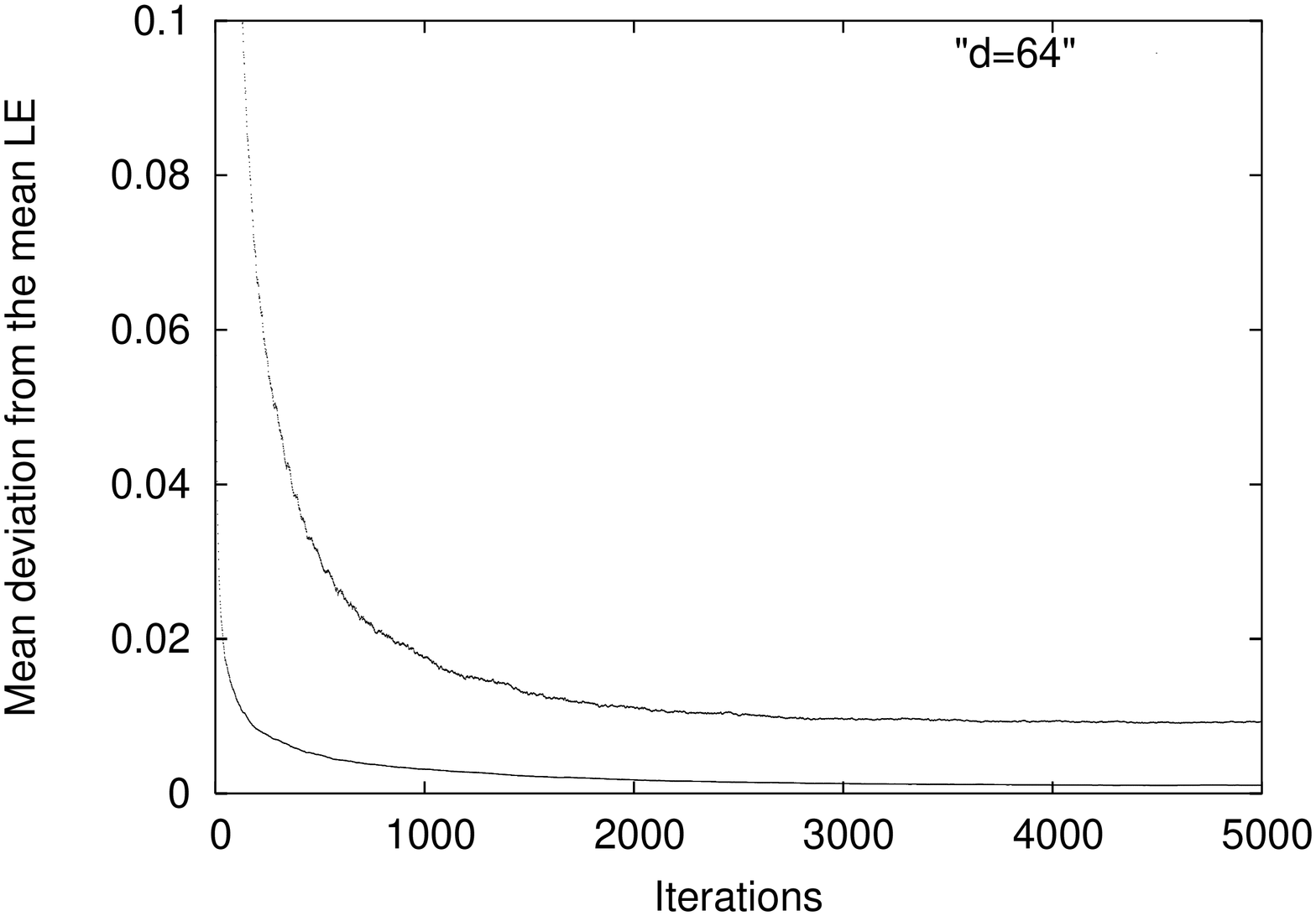, height=2.3in, angle=270}
\caption{Mean deviation from the mean of the largest and most negative
  Lyapunov exponent per time-step
  for an ensemble of $1000$ networks with $32$
  neurons and $16$ (left) and $64$ (right)  dimensions}
\label{fig:meanlceerror16and64}
\end{center}
\end{figure*}

Figures (\ref{fig:lcevsiteration16and64}) and (\ref{fig:closeup64})
provide insight into how the individual exponents for individual
networks converge; we now must establish the convergence of the
Lyapunov exponents for a large set of neural networks and present
a general idea of the numerical variance ($\epsilon_m$) in the
Lyapunov exponents.  We will achieve this in the following manner: we
will calculate the Lyapunov spectrum for an individual network for
$5000$ time steps; we will calculate the mean of each exponent in the
spectrum; we will, for each time step calculate the deviation of the
exponent from the mean of that exponent; we will follow the above
procedure for $1000$ networks and take the mean of the deviation from
the mean exponent at each time step.  Figure
(\ref{fig:meanlceerror16and64}) represents the analysis in the former
statement.  This figure demonstrates clearly that the deviation from the mean exponent, even
for the most negative exponent (the most negative exponent
has the largest error) drops below $0.01$ after $3000$ time
steps.  The fluctuations in the largest Lyapunov exponent lie in the
$10^{-3}$ range for $3000$ time-steps.  Figure (\ref{fig:meanlceerror16and64}) also substantiates three notions: a measurement of how little the average exponent strays
from its mean value; a measurement of the similarity of this
characteristic over the ensemble of networks; and finally it helps
establish a general intuition with respect to the accuracy of our
exponents, $\epsilon_m < 0.01$ for $5000$ time steps.  

It is worth noting that determining errors in the Lyapunov exponents
is not an exact science; for our networks such errors vary a great
deal in different regions in $s$ space.  For instance, near the first
bifurcation from a fixed point can require up to $100000$ or more
iterations to converge to an attractor and $50000$ more iterations for
the Lyapunov spectrum to converge.

\section{Numerical arguments for preliminaries}
\label{sec:prelimnarguments}
Before we present our arguments supporting our conjectures we must
present various preliminary results.  Specifically we will discuss the $num-$continuity
of the Lyapuonv exponents, the $a-$density of Lyapunov exponent
zero-crossings, and argue for the existence of arbitrarily high number
of positive exponents given an arbitrarily high number of dimensions.  With these preliminaries in place, the arguments
supporting  our conjectures will be far more clear.

\subsection{$num-$continuity}
\label{sec:ncontinuity}
Testing for the $num-$continuity of Lyapunov exponents formally will be
two-fold.  First, we will need to investigate,
for a specific network, $f$, the behavior of Lyapunov exponents versus
variation of parameters.  Second, indirect, yet strong evidence of
the $num-$continuity will also come from investigating how periodic
window size varies with dimension and parameter variation.  It is
important to note that when we refer to continuity, we are referring
to a very local notion of continuity.  Continuity is always in
reference to the set upon which something (a function, a mapping,
etc) is continuous.  In the below analysis, the neighborhoods upon
which continuity of the Lyapunov exponents are examined is over ranges
of plus and minus one parameter increment.  This is all that is
necessary for our purposes, but this analysis cannot guarantee
strict continuity along, say, $s \in [0.1, 10]$, but rather continuity
along little linked bits of the interval $[0.1, 10]$. 

\subsubsection{Qualitative analysis}
\begin{figure*}
\begin{center}
\epsfig{file=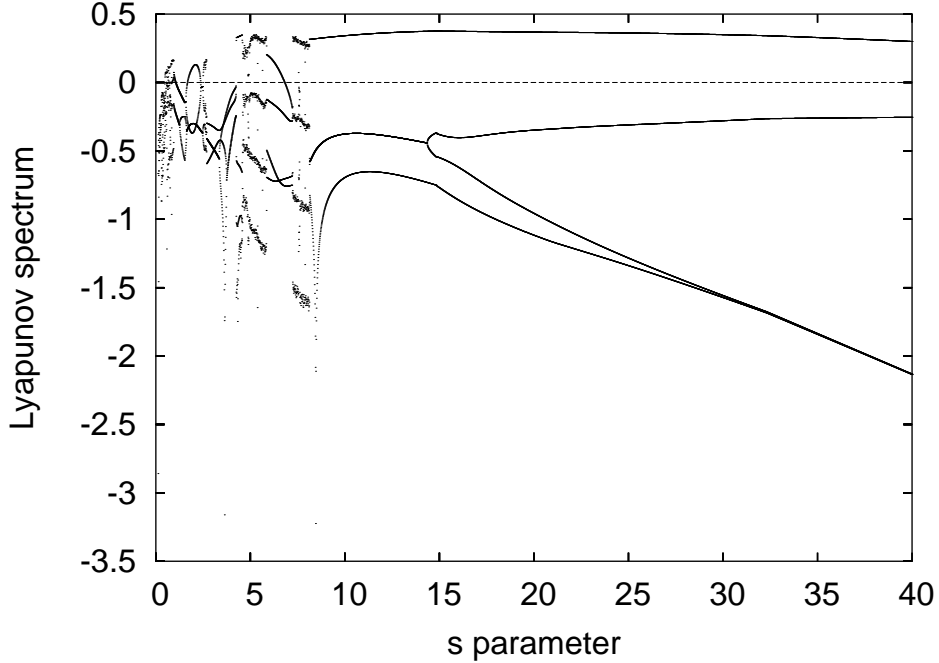, height=13cm, angle=270}
\caption{LE spectrum: $32$ neurons, $4$ dimensions.}
\label{fig:biglced4}
\end{center}
\end{figure*}

\begin{figure*}
\begin{center}
\epsfig{file=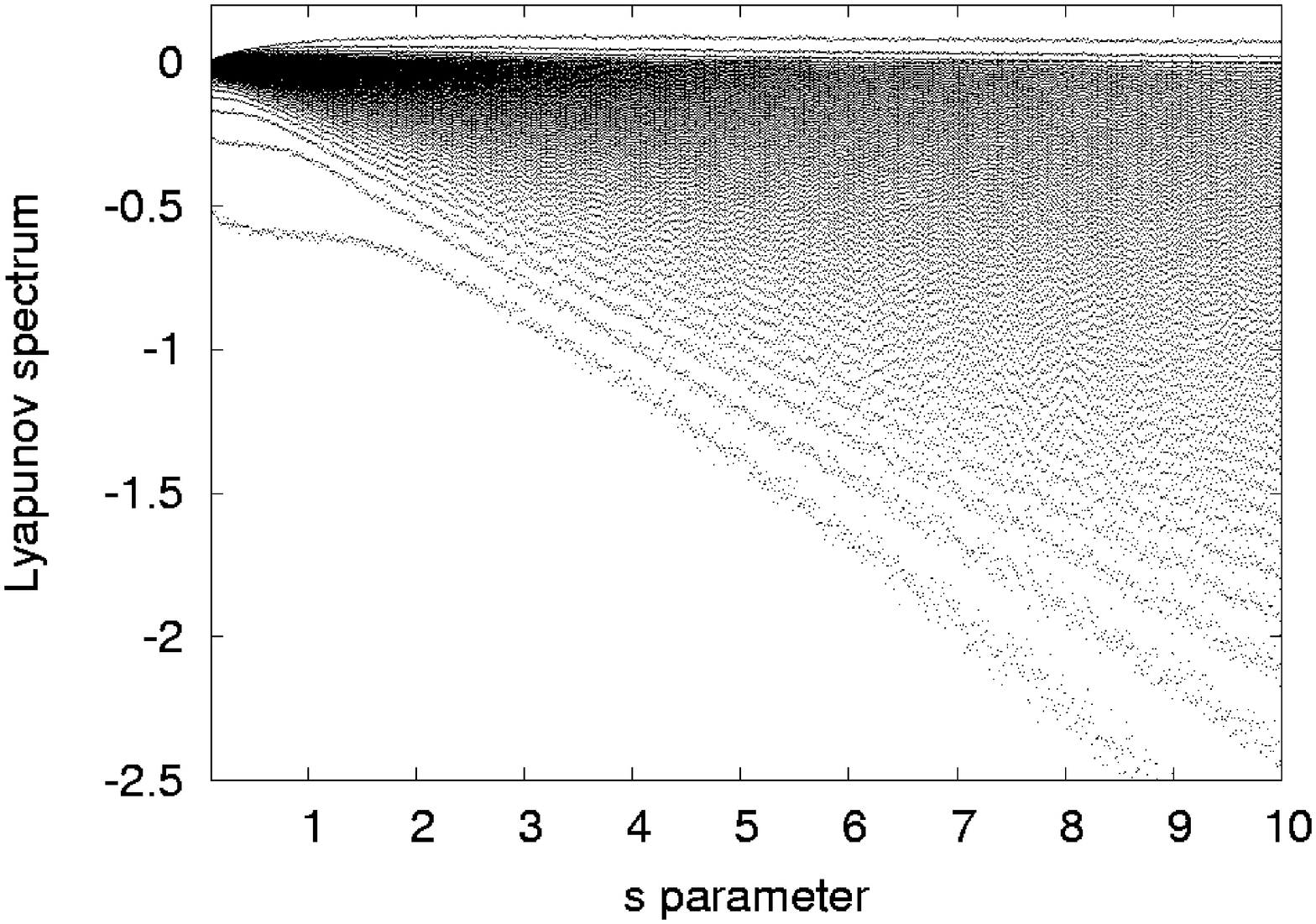, height=9cm, angle=270}
\caption{LE spectrum: 32 neurons, 64 dimensions.}
\label{fig:biglced64}
\end{center}
\end{figure*}

Qualitatively, our intuition for $num-$continuity comes from examining hundreds of Lyapunov spectrum plots versus parameter variation.  In
this vein, Figs. (\ref{fig:biglced4}) and (\ref{fig:biglced64}) present the difference between low
and higher dimensional Lyapunov spectra.

In Fig. (\ref{fig:biglced64}), the Lyapunov exponents look continuous within
numerical errors (usually $\pm 0.005$).  Figure (\ref{fig:biglced64}) by itself provides little more
than an intuitive picture of what we are attempting to argue.  As we
will be making arguments that the Lyapunov spectrum will become more
smooth, periodic windows will disappear, etc, with increasing
dimension, Fig. (\ref{fig:biglced4}) shows a typical graph of the Lyapunov
spectrum versus parameter variation for a neural network with $32$
neurons and $4$ dimensions.  The contrast between Figs. (\ref{fig:biglced64}) and (\ref{fig:biglced4}) intuitively demonstrates the
increase in continuity we are claiming. 

Although a consideration of Figs. (\ref{fig:biglced4}) and
(\ref{fig:biglced64}) yields an observation that, as the dimension is
increased, the Lyapunov exponents appear to be more continuous
function of the $s$ parameter, the above figures alone do not verify
$num-$continuity.  In fact, it should be noted that pathological
discontinuities have been observed in networks with as many as $32$ dimensions.  The
existence of pathologies for higher dimensions is not a problem we are
prepared to answer in depth; it can be confidently said that as the
dimension (number of inputs) is increased, the frequency of
pathologies appears to become vanishingly rare (this is noted over our
observation of several thousand networks with dimensions ranging from
$4$ to $256$). 

\subsubsection{Quantitative and numerical analysis}
Our quantitative analysis will follow two lines.  The first will be
a specific analysis along the region of parameter change for three
networks with dimensions $4$ and $64$, respectively.  This will be
followed with a more statistical study of a number of networks per
dimension where the dimensions will range from 4 to 128 in powers of 2.

\begin{figure}
\begin{center}
\epsfig{file=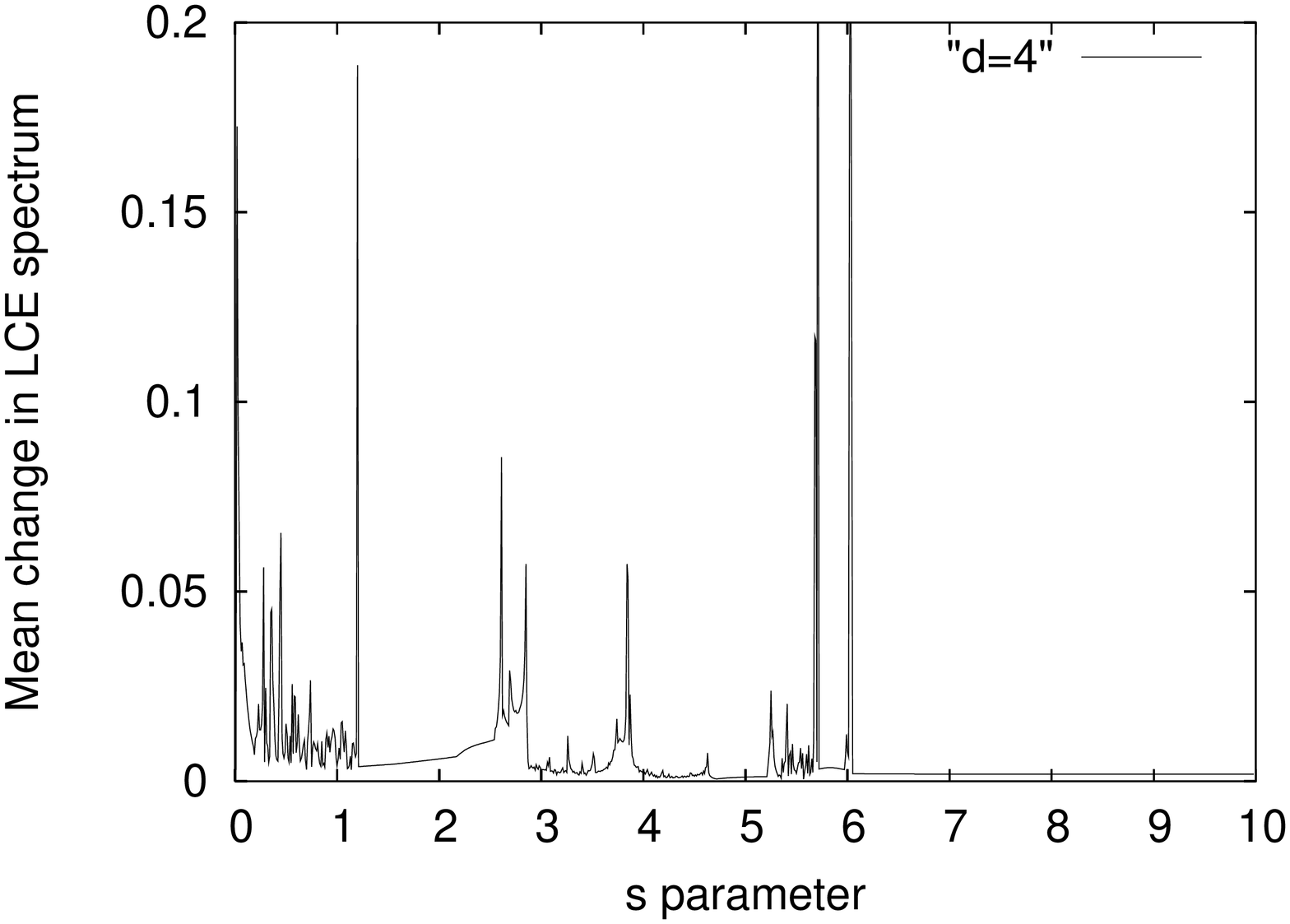, height=7cm, angle=270}
\epsfig{file=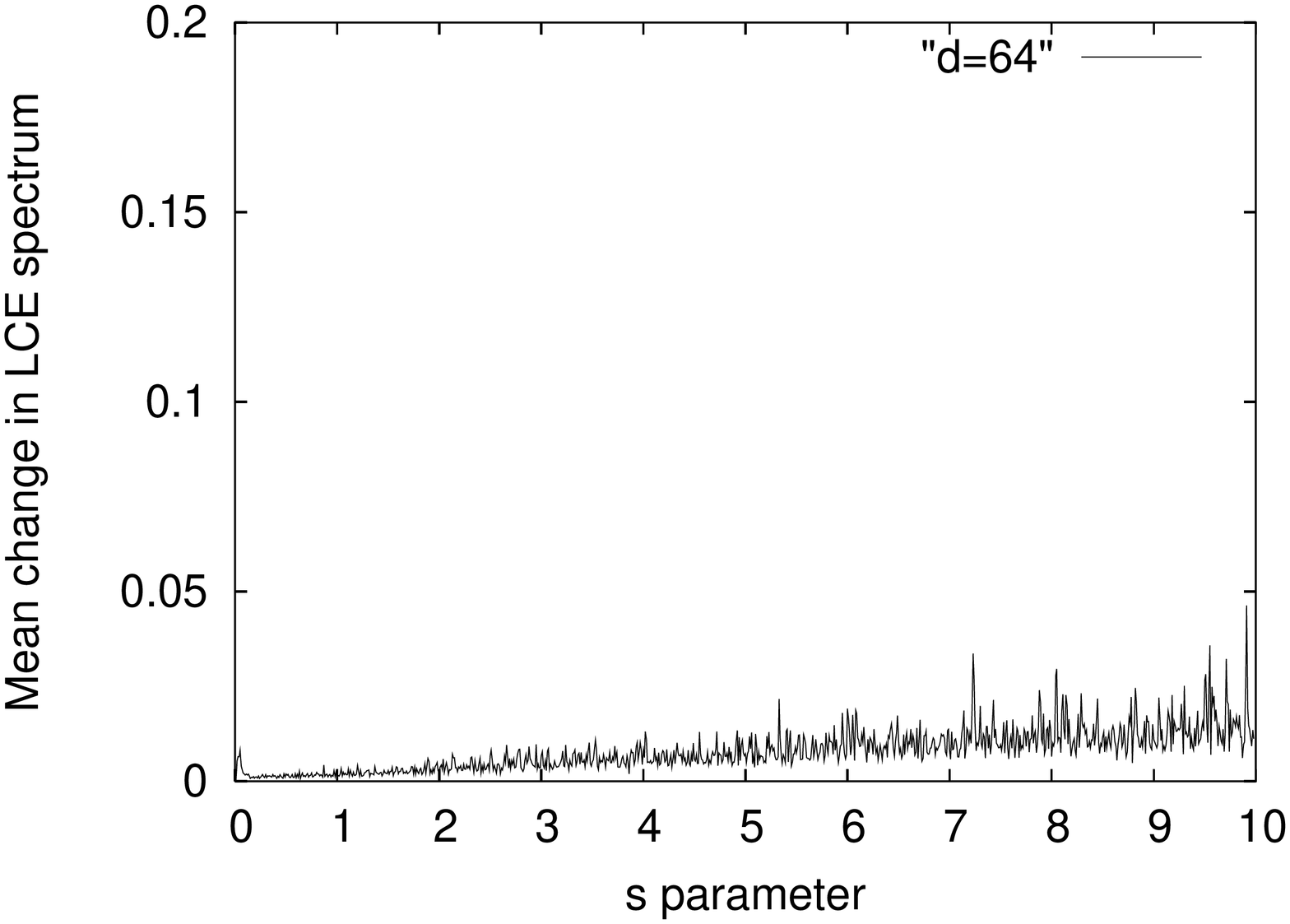, height=7cm, angle=270}
\caption{$num-$continuity (mean of $|\chi_{i}(s) - \chi_i(s + \delta
  s)|$ for each $i$) versus parameter variation: 32 neurons, 4 (left) and 64 (right) dimensions.}
\label{fig:ncpvariation}
\end{center}
\end{figure}

Consider the $num-$continuity of two different networks
while varying the $s$ parameter.  Figure (\ref{fig:ncpvariation}) is a
plot of the mean
difference in each exponent between parameter values summed over all the exponents.
The parameter increment is $\delta s = 0.01$. 

The region of particular interest is between $s=0$ and $6$.  Considering
this range, it is clear that the variation in the
mean of the exponents versus variation in $s$ decreases with
dimension.  The $4$-dimensional network not only has a higher baseline
of $num-$continuity, but it also has many large spikes.  As the
dimension is increased, considering the $64$-dimensional case,
the baseline of $num-$continuity is decreased, and the large spikes
disappear.  The spikes in the $4$-dimensional case can be directly linked
to the existence of periodic windows and bifurcations that result in
dramatic topological change.  This is one verification of
$num-$continuity of Lyapunov exponents.  These two cases are quite typical, but it is clear that
the above analysis, although quite persuasive, is not adequate for our
needs.  We will thus resort to a statistical study of the above plots.

\begin{figure}
\begin{center}
\epsfig{file=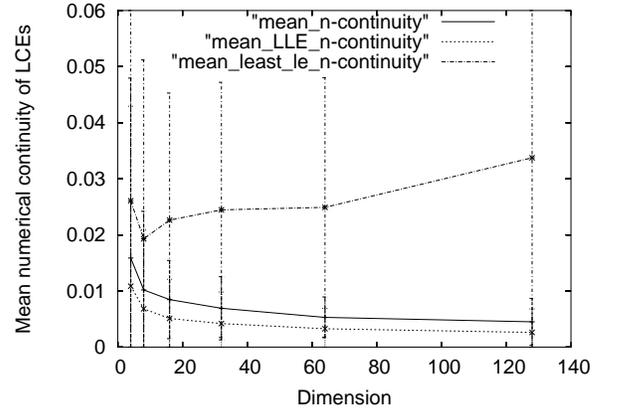, height=8cm, angle=270}
\caption{Mean $num-$continuity, $num-$continuity of the largest and the
  most negative Lyapunov exponent of many networks versus their dimension.  The error bars are the standard
  deviation about the mean over the number of networks considered.}
\label{fig:ncversusd}
\end{center}
\end{figure}

The statistical support we have for our claim of increased
$num-$continuity will focus on the parameter region between $s=0$ and
$6$, the region in parameter space over which the maxima of entropy,
Kaplan-Yorke dimension, and the number of positive Lyapunov exponents exists.
Figure (\ref{fig:ncversusd}) considers the $num-$continuity along parameter values
ranging from $0$ to $6$.  The points on the plot correspond to the
mean (over a few hundred networks) of the mean exponent change between
parameter values, or:
\begin{equation}
\mu^d = \frac{1}{Z} \sum_{k=1}^{Z} \frac{\sum_{i=1}^{d}
  |\chi_i^k (s) - \chi_i^k(s
+ \delta s)|  }{d}  
\end{equation}
where $Z$ is the total number of networks of a given dimension considered.

Figure (\ref{fig:ncversusd}) clearly shows that as the dimension is increased, for
the same computation time, both the mean exponent change versus
parameter variation per network
and the standard deviation of the exponent change decrease
substantially as the dimension is increased.\footnote{The mean
  $num-$continuity for $d=4$ and $d=128$ is $0.015 \pm 0.03$ and $0.004
  \pm 0.003$, respectively.  The mean
  $num-$continuity of the largest exponent for $d=4$ and $d=128$ is
  $0.01 \pm 0.03$ and $0.002 \pm 0.004$, respectively.  The discrepancy
between these two data points comes from the large error in the
negative exponents at high dimension.}  Of course the mean
change over all the exponents allows for the possibility for one
exponent (possibly the largest exponent) to undergo a relatively large
change while the other exponents change very little.  For this
reason, we have included the $num-$continuity of the largest and the
most negative exponents versus parameter change.  The $num-$continuity
of the largest exponents is very good, displaying a small standard
deviation across many networks.  The error in the most negative
exponent is inherent to our numerical techniques (specifically the
Gram-Schmidt orthogonalization).  The error in the most negative
exponent increases with dimension, but is a numerical artifact.  This
figure yields strong
evidence that in the region of parameter space where the network starts at a fixed point
(all negative Lyapunov exponents), grows to having the maximum number of positive
exponents, and returns to having a few positive exponents, the
variation in any specific Lyapunov exponent is very small.  

\begin{figure}
\begin{center}
\epsfig{file=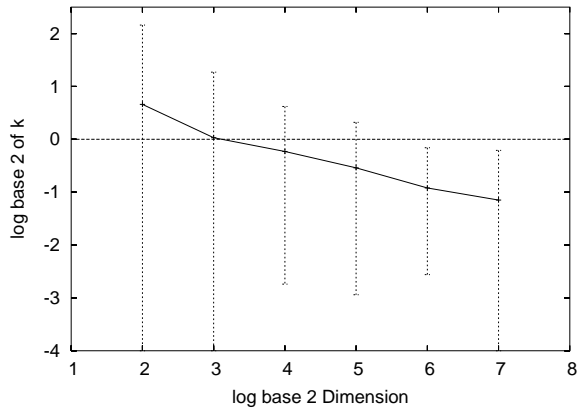, height=8cm, angle=270}
\caption{$k$-scaling: $\log_2$ of dimension versus $\log_2$ of
  $num-$Lipschitz constant of the largest Lyapunov exponent.}
\label{fig:kscaling}
\end{center}
\end{figure}

There is a specific relation between the above data to definition
\ref{definition:nlipschitz}; $num-$Lipschitz is a stronger condition
than $num-$continuity of Lyapunov exponents.  The mean $num-$continuity at $n=32$, $d=4$
\begin{align}
|\chi_j(s+\delta_{num}) - \chi_j(s)| < k \delta_{num} \\
|0.02| < k |0.01|
\end{align} 
yielding $k=2$ which would not classify as $num-$Lipschitz contracting, whereas
for $n=32$, $d=128$ we arrive at
\begin{align}
|\chi_j(s+\delta_{num}) - \chi_j(s)| < k \delta_{num} \\
|0.004| < k |0.01|
\end{align}
which yields $k=0.4<1$ which does satisfy the condition for $num-$Lipschitz contraction.  Even more striking is the $num-$continuity of
only the largest Lyapunov exponent; for $n=32$, $d=4$ we get
\begin{align}
|\chi_j(s+\delta_{num}) - \chi_j(s)| < k \delta_{num} \\
|0.015| < k |0.01|
\end{align}
which yields $k=1.5$, while the $n=32$ $d=128$ case is
\begin{align}
|\chi_j(s+\delta_{num}) - \chi_j(s)| < k \delta_{num} \\
|0.002| < k |0.01|
\end{align}
which nets $k=0.2$.  As the dimension is increased, $k$ decreases,
and thus $num-$continuity increases.  As can be seen from Fig. (\ref{fig:ncversusd}),
the $num-$continuity is achieved rather quickly as the dimension is
increased; the Lyapunov exponents are quite continuous with respect to
parameter variation by $16$ dimensions.  For an understanding in an
asymptotic limit of high dimension, consider Fig. (\ref{fig:kscaling}).  As the dimension is increased the $\log_2$ of
the dimension versus the $\log_2(k_{\chi_1})$ yields the scaling $k
\sim \sqrt(\frac{2}{d})$; thus as $d \rightarrow \infty$, $k_{\chi_1}
\rightarrow 0$, which is exactly what we desire for continuity in the
Lyapunov exponents versus parameter change.  This completes our evidence
for the $num-$continuity in high-dimensional networks. 

\subsubsection{Relevance}

Conjectures (\ref{conjecture:hypvol}), (\ref{conjecture:zerobifvol}),
and (\ref{conjecture:nongenercityss}) are all fundamentally based on
condition (\ref{condition:ass2}).  For the
neural networks, all we need to establish conjecture
(\ref{conjecture:hypvol}) is the $num-$continuity of the Lyapunov exponents,
the existence of the fixed point for $s$ near $0$, the periodic orbits
for $s \rightarrow \infty$, and  three exponents that are, over some
region of parameter space, all simultaneously positive.  The n-continuity of Lyapunov exponents implies, within
numerical precision, that Lyapunov exponents both pass through zero (and
don't jump from positive to negative without passing through zero) and
are, within numerical precision, zero.

\subsection{$a-$density of zero crossings}
\label{sec:adensity}
Many of our arguments will revolve around varying $s$ in a range of
$0.1$ to $6$ and studying the behavior of the Lyapunov spectrum.  One of
the most important features of the Lyapunov spectrum we will need is
a uniformity in the distribution of positive exponents between $0$ and
$\chi_{max}$.  As we are dealing with a countable set, we will refrain from
any type of measure theoretic notions, and instead rely on $a$-density of
the set of positive exponents as the dimension is increased.  Recall
the definition of $a$-dense (definition
(\ref{definition:eventuallydense})), the definition of a
bifurcation chain subset (definition
(\ref{definition:bifurcationchainsubset})), which corresponds to the
set of Lyapunov exponent zero crossings, and the definition of a chain
link set (definition (\ref{definition:chainlinkset})).  Our conjectures will make sense if and only if, as the dimension is increased, the bifurcation chain subsets become
``increasingly'' dense, or $a$-dense in the closure of the chain link
set ($\bar{V}$).  The notion of $a$-dense bifurcation chain set in the
closure of the chain link set as dimension is increased that provides us with the
convergence to density of non-hyperbolic points we need to
satisfy our goals.

\subsubsection{Qualitative analysis}
The qualitative analysis will focus on pointing out what
characteristics we are
looking for and why we believe $a-$density of Lyapunov exponent
zero-crossings ($a$-dense bifurcation chain set in the
closure of the chain link set)
over a particular region of parameter space exists.  A byproduct of this
analysis will be a picture of one of the key traits needed to support
our conjectures.  We will begin with figures showing the positive Lyapunov
spectrum for $16$ and $64$ dimensions. 
 
\begin{figure}
\begin{center}
\epsfig{file=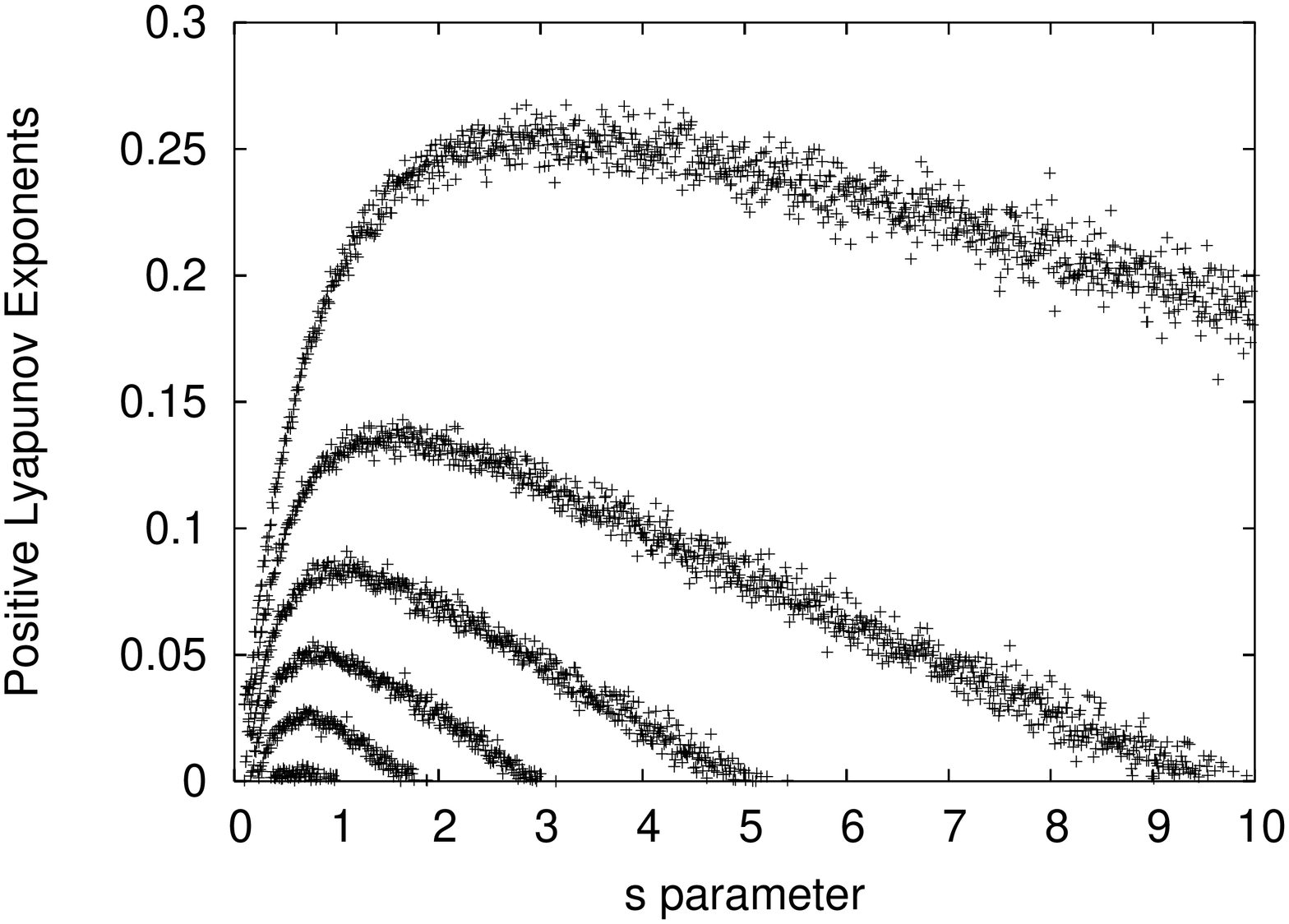, height=7cm, angle=270}
\epsfig{file=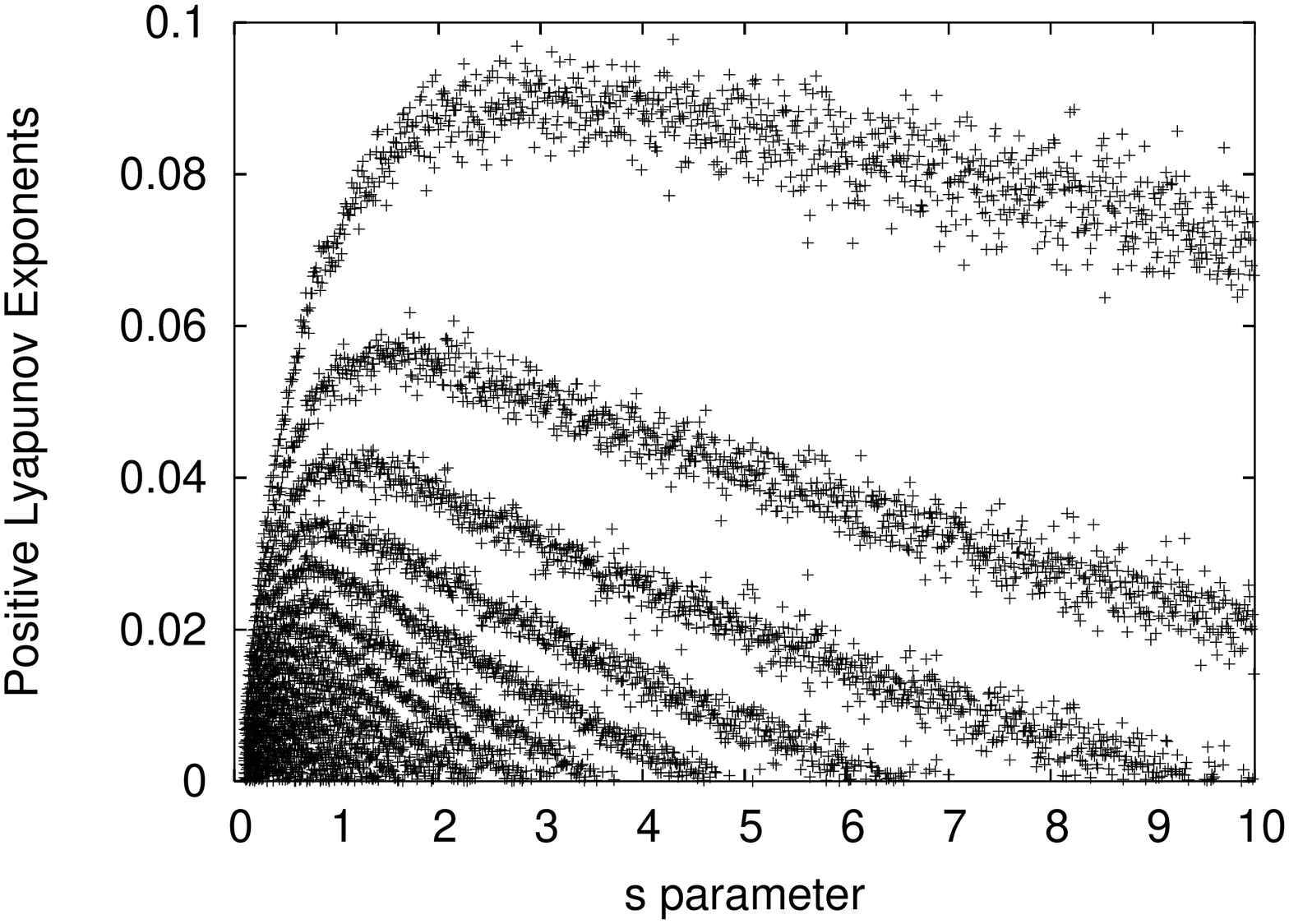, height=7cm, angle=270}
\caption{Positive LE spectrum for typical individual networks with $32$
  neurons and $16$ (top) and $64$ (bottom) dimensions.}
\label{fig:poslce4and64}
\end{center}
\end{figure}

Considering the $16$-dimensional case, and
splitting the $s$ parameter variation into two regions, region one - $R_I = [0,
0.5]$, and region two - $R_{II} = [0.5, 10]$.  We then partition up
$R_{II}$ using the bifurcation link sets, and collect the zero
crossings in the bifurcation chain sets.    

We want the elements of the bifurcation chain sets to be spaced evenly enough so that,
as the dimension goes to infinity, variations in the $s$ parameter on
the chain link set will lead to a Lyapunov exponent zero-crossing (and a transition from $V_i$ to $V_{i \pm 1}$)\footnote{Recall,
  the bifurcation chain sets will not exist when the zero crossings
  are not transverse.}.  Considering region $II$\footnote{We will save region $I$ for a
  different report.  For insight into some of the dynamics and
  phenomena of region $I$, see \cite{albersroutetochaosI}.}, we wish for the distance along the $s$
axis between Lyapunov exponent zero-crossings (elements of the bifurcation
chain subset) to decrease as the dimension
is increased.  If, as the dimension is increased, the Lyapunov
exponents begin to ``bunch-up'' and cease to be at least somewhat uniformly distributed,
the rest of our arguments will surely fail.  For
instance, in region two of the bottom plot of Fig. (\ref{fig:poslce4and64}), if the Lyapunov exponents were
``clumped,'' there will be many holes where variation of
$s$ will not imply an exponent crossing.  Luckily,
considering the $64$-dimensional case as given in
Fig. (\ref{fig:poslce4and64}), our desires seem to be as the spacing between exponent zero-crossings is clearly decreasing as the
dimension is increased (consider
the region $[0.5, 4]$), and there are no point accumulations of
exponents.  It is also reassuring to note that even at $16$ dimensions,
and especially at $64$ dimensions, the Lyapunov exponents are quite
distinct and look $num-$continuous as previously asserted.  The above
figures are, of course, only a picture of two networks; if we wish for
a more conclusive statement, we will need arguments of a statistical nature.

\subsubsection{Quantitative and numerical analysis}
Our analysis that specifically targets the $a-$density of Lyapunov
exponent zero crossings focuses on an analysis of plots of the number
of positive exponents versus the $s$ parameter. 

\begin{figure}
\begin{center}
\epsfig{file=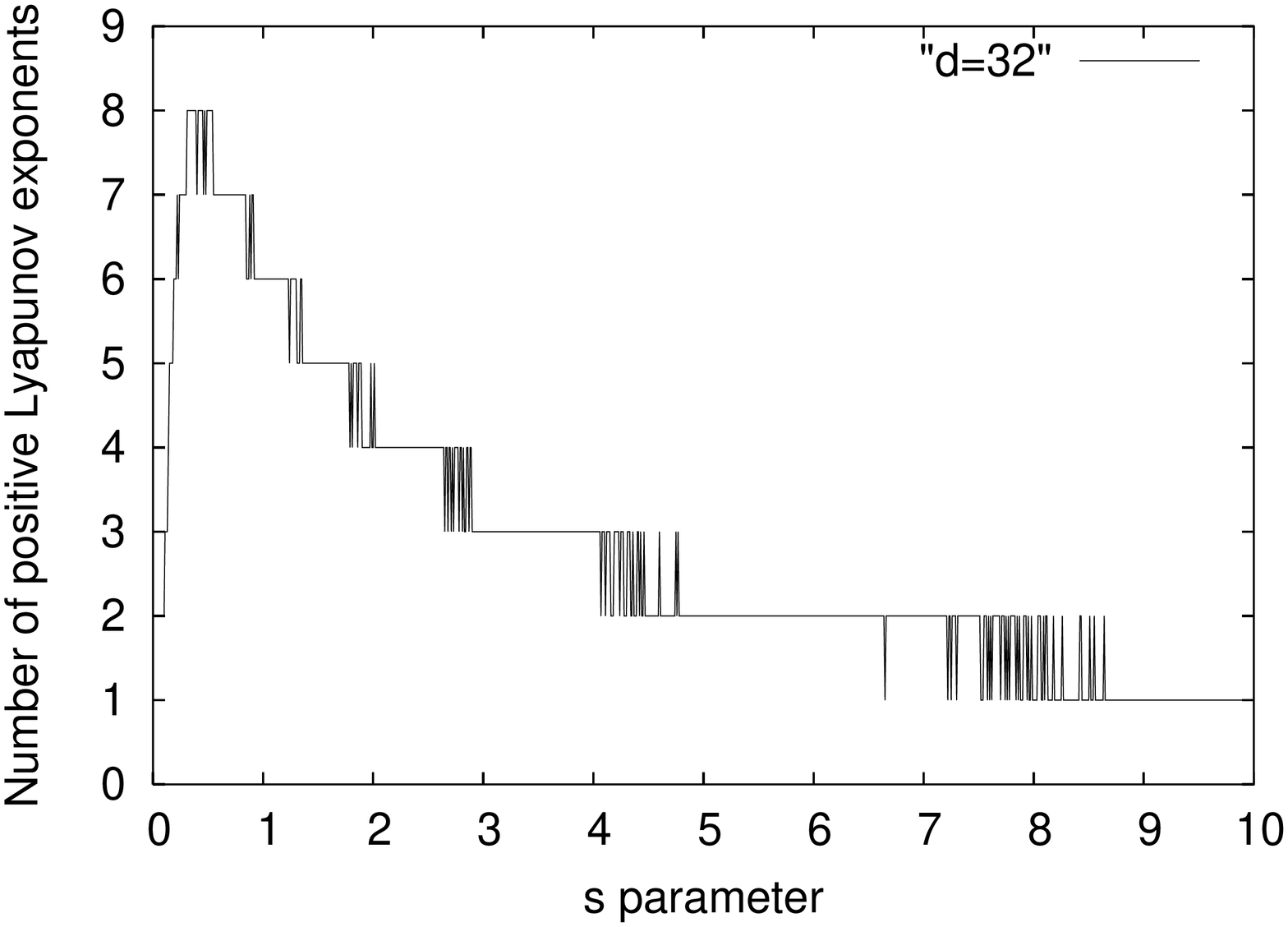, height=7cm, angle=270}
\epsfig{file=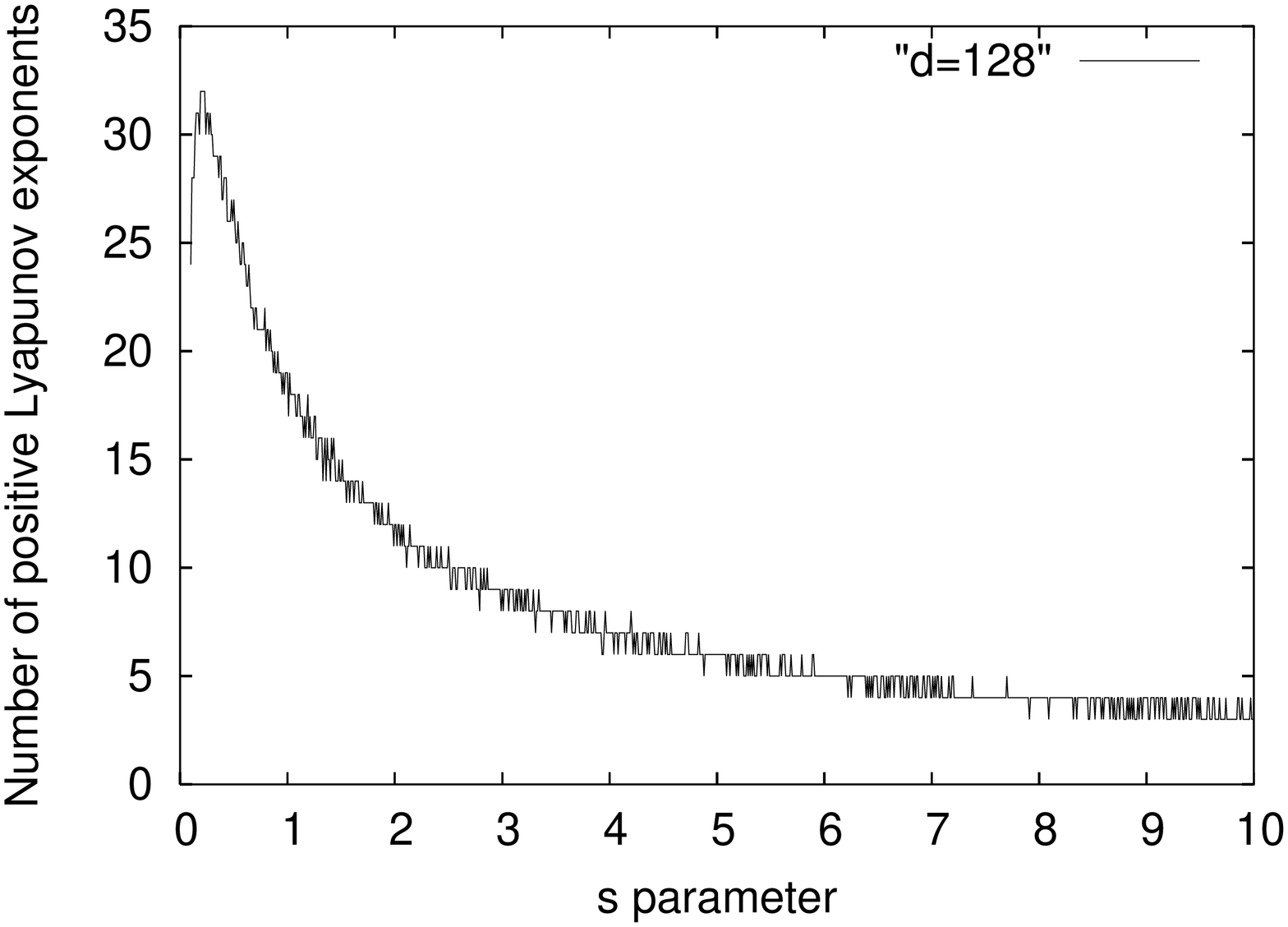, height=7cm, angle=270}
\caption{Number of positive LE's for typical individual networks with $32$
  neurons and $32$ (top) and $128$ (bottom) dimensions.}
\label{fig:numberoflcepos}
\end{center}
\end{figure}

Qualitatively, the two examples given in Fig. (\ref{fig:numberoflcepos}) (both of which typify the
behavior for their respective number of neurons and dimensions) exemplify
the $a-$density for which we are searching.  As the dimension is
increased, the plot of the variation in the number of positive
exponents versus $s$ becomes more smooth\footnote{This increase in
  smoothness is not necessarily a function of an increased number of
  exponents.  A dynamical system that undergoes
  massive topological changes upon parameter variation will not have a
  smooth curve such as in Fig. (\ref{fig:numberoflcepos}), regardless
  of the number of exponents.}, while the width of the
peak becomes more narrow.  Thus, the slope of the number of
positive exponents versus $s$ between $s=s_{*}$ ($s_{*}$
  is $s$ where there exists the maximum number of positive Lyapunov
  exponents), and $s=2$ drops from $-3$ at $d=32$ to $-13$ at $d=128$.  Noting that
the more negative the slope, the less varition in $s$ is required to
force a zero-crossing, it is clear that this implies $a-$density of
zero-crossings.  We will not take that line of analysis
further, but rather will give brute force evidence for $a-$density by directly noting the mean distance between
exponent zero-crossings.

\begin{center}
\begin{figure}
\epsfig{file=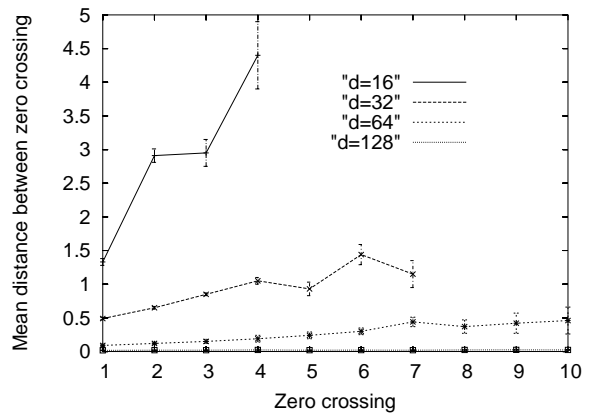, height=8cm, angle=270}
\caption{Mean distance between the first $10$ zero crossings of LE's
  for many networks with $32$ neurons and $16$, $32$, $64$, and $128$ dimensions.}
\label{fig:adensity}
\end{figure}
\end{center}

From Fig. (\ref{fig:adensity}), it is clear that as the dimension of the
network is increased, the mean distance between successive exponent
zero-crossings decreases.  Note that measuring the mean distance
between successive zero-crossings both in an intuitive and brute force
manner, verifies  the sufficient condition for
the $a-$density of the set of $s$ values for which there exist zero-crossings of exponents.  The error bars represent the standard
deviation of the length between zero-crossing over an ensemble
(several hundred for low dimensions, on the order of a hundred for
$d=128$) networks.  For the cases where the dimension was $16$ and $32$, the
$s$ increment resolution was $\delta s = 0.01$.  The error in the zero
crossing distance for these cases is, at the smallest, $0.02$, while
at its smallest the zero crossing distance is $0.49$, thus
resolution of $0.01$ in $s$ variation is sufficient to adequately
resolve the zero crossings.  Such is not the case for $64$ and $128$
dimensional networks.  For these cases we were required to increase
the $s$ resolution to $0.005$.  The zero crossings of a few hundred networks considered
were all examined by hand; the distances between the zero crossings were
always distinct, with a resolution well below that necessary to determine
the zero crossing point.  The errors were also determined by hand,
noting the greatest, and least reasonable point for the zero
crossing.  All the zero crossings were determined after the
smallest positive exponent that became positive hit its peak value,
i.e. after approximately $0.75$ in the $d=16$ case of Fig. (\ref{fig:poslce4and64}).


\subsubsection{Relevance}
The $a-$density of zero crossings of Lyapunov exponents provides
the most important element in our arguments of conjectures
(\ref{conjecture:hypvol}) and (\ref{conjecture:zerobifvol}); combining
$num-$continuity with $a-$density will essentially net our desired
results.  If continuity of Lyapunov exponents increases, and the
density of zero crossings of exponents increases over a set $U \in
R^1$ of parameter space, it seems clear that we will have both
hyperbolicity violation and, upon variation of parameters in $U$, we
will have the topological change we are claiming.  Of course small issues remain, but those will be dealt with
in the final arguments.

\subsection{Arbitrarily large number of positive exponents}
\label{sec:arbnumberofposexponents}
For our $a-$density arguments to work, we need a set whose cardinality
is asymptotically a countably infinite
set (such that it can be $a-$dense in itself) and we need the distance
between the elements in the set to approach zero.  The later
characteristic was the subject of the previous section, the former
subject is what we intend to address in this section.

\subsubsection{Qualitative analysis}
The qualitative analysis of this can be seen in Fig. (\ref{fig:numberoflcepos}); as the dimension is increased, the maximum
number of positive Lyapunov exponents clearly increases.  We wish to
quantify that the increase in the number of positive exponents
versus dimension occurs for a statistically relevant set of networks.
 
\subsubsection{Quantitative analysis}

\begin{figure}
\begin{center}
\epsfig{file=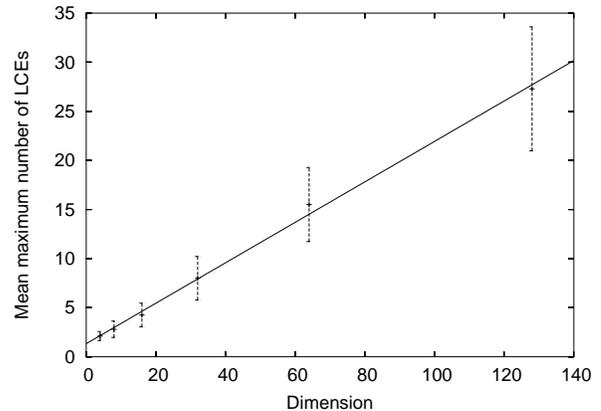, height=8cm, angle=270}
\caption{Mean maximum number of positive LE's versus dimension,
  all networks have $32$ neurons (slope is approximately $\frac{1}{4}$).}
\label{fig:maxnumberposlcevsd}
\end{center}
\end{figure}

We will use a brute force argument to demonstrate the increase in
positive Lyapunov exponents with dimension; we will simply
plot the number of positive exponents at the maximum
number of exponents as dimension is increased.  We claim that the number of Lyapunov exponents increases and, in fact,
diverges to infinity as the limit dimension of the network is taken to
infinity.  Figure (\ref{fig:maxnumberposlcevsd}) showing the number of positive
Lyapunov exponents versus dimension.

From Fig. (\ref{fig:maxnumberposlcevsd}) it is clear that as the dimension is increased, the
number of positive exponents increases in a nearly linear
fashion \footnote{Further evidence for such an increase is provided by
  considering the Kaplan-Yorke dimension versus $d$.  Such analysis
  yields a linear dependence,  $D_{K-Y} \sim d/2$.
}.  Further, this plot is linear to as high a dimension as the
authors could compute enough cases for reasonable statistics.  If the
maximum number of exponents versus dimension remains linear beyond
the range we could compute, we will have the countably infinite number of
positive exponents we require. 

\subsubsection{Relevance}
The importance of the increasing number of positive exponents with
dimension is quite simple.  For the $a-$density of exponent zero
crossing to be meaningful in the infinite-dimensional limit, there
must also be an arbitrarily large number of positive exponents that
can cross zero.  If, asymptotically, there is a finite number of
positive exponents, all of our claims will be false; $a-$density
requires a countably infinity set.

\section{Numerical arguments for conjectures}
\label{sec:nargumentsofconjectures}

\subsection{Decreasing window probability}
\label{sec:decreasingwindowsize}
With the $num-$continuity and $a-$density arguments already in place,
all the evidence required to show the length of periodic windows along a curve in parameter
space is already in place.  We will present a bit of new
data, but primarily we will clarify exactly what the conjecture says.
We will also list the specifics under which the conjecture applies in our circumstances.  

\subsubsection{Qualitative analysis}
Qualitative evidence for the dissappearance of periodic windows amidst chaos is evident from
Figs. (\ref{fig:biglced4}), (\ref{fig:biglced64}) and
(\ref{fig:poslce4and64}); the periodic windows that dominate the
$4$-dimensional network over the parameter range $s=0$ to $10$
are totally absent in the $64$-dimensional
network.  It is important to note that for this conjecture, as well as
all our conjectures, we are considering the $s$ parameter over ranges
no larger than $0$ to $10$.  We will avoid, for the most part, the
``route to chaos'' region ($s$ near zero), as it yields many complex issues that will
be saved for another report.  We will instead consider the parameter
region after the lowest positive exponent first becomes positive.  We could consider parameter ranges
considerably larger, but for $s$ very large, the round-off error
begins to play a significant role, and the networks become binary.
This region has been briefly explored in \cite{albersroutetochaosI}; further analysis
is necessary for a more complete understanding \cite{intermittper}.  

\subsubsection{Quantitative and numerical analysis}
The quantitative analysis we wish to perform will involve arguments
of two types; those that are derived from data given in
sections (\ref{sec:ncontinuity}) and (\ref{sec:adensity}), and those that follow from statistical data
regarding the probability of a window existing for a given $s$ along
an interval in $R$.  We
begin by recalling what we are attempting to claim and what conditions
we need to verify the claim.  We will then present the former argument and conclude with the latter.

The conjecture we are investigating claims that as the dimension of a
dynamical system is increased, periodic windows
along a one-dimensional curve in parameter space vanish in a significant portion of parameter space
for which the dynamical system is chaotic.  This is, of
course, dependent upon the region of parameter space one is
observing --- and there is likely no way to rid ourselves of such an
issue.  For our purposes, we will generally be investigating the region of
$s$ parameter space between $0.1$ and $10$, however, sometimes we will
limit the investigation to s between $2$ and $4$.  Little changes if we
increase $s$ until the network begins behaving as a binary system
due (quite possibly) to the round-off error.  However, along the
transition to the binary region, there are significant complications
which we will not address here.  As the dimension is increased, the main concern is that the lengths of the bifurcation chain sets
must increase such that there will exist at least one bifurcation
chain set that has a cardinality approaching infinity as the dimension of the network approaches
infinity.

Our first argument is based directly upon the evidence of
$num-$continuity of Lyapunov exponents.  From Fig. (\ref{fig:ncversusd}) it is
clear that as the dimension of the set of networks sampled is increased,
the mean difference in Lyapunov exponents over small ($\delta s
=0.01$) $s$ parameter perturbation decreases.  This increase in
$num-$continuity of the Lyapunov exponents with dimension over our
parameter range is a direct
result of the disappearance of periodic windows from the chaotic
regions of parameter space.  This evidence is amplified by the
decrease in the standard deviation of the $num-$continuity versus
dimension (of both the
mean of the exponents and the largest exponent).  This decrease in the
standard deviation of the $num-$continuity of the largest Lyapunov
exponent allows for
the existence of fewer large deviations in Lyapunov exponents (large deviations
are needed for all the exponents to suddenly become less than or equal
to zero).  

\begin{figure}
\begin{center}
\epsfig{file=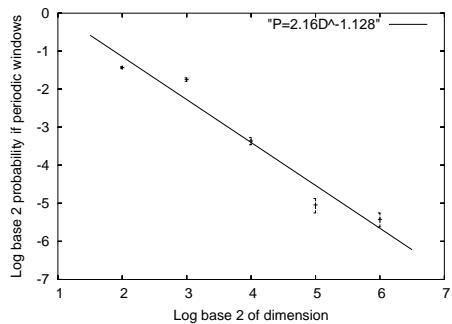, height=6.2cm, angle=270}
\caption{$\log_2$ of the probability of periodic or quasi-periodic
  windows versus $\log_2$ of dimension.  The line $P_w = 2.16
  d^{-1.128}$ is the least squares fit of the plotted data.}
\label{fig:probofwindows}
\end{center}
\end{figure}

We can take this analysis a step further and simply calculate the
probability of an $s$ value having a periodic orbit over a given
interval.  Figure (\ref{fig:probofwindows}) shows the probability
of a periodic window existing for a given $s$ on the interval $(2, 4)$ with
$\delta s = 0.001$ for various dimensions.  There is a power law in
the probability of periodic windows --- the probability of the
existence of a period window decreases approximately as $ \sim \frac{1}{d}$.  Moreover, the authors have observed that in high
dimensional dynamical systems, when periodic windows are observed on
the interval $(2, 4)$, they are usually large in length.  In other words, even
though the probability that a given $s$ value will yield a periodic
orbit for $d=64$ is $0.02$, it is likely that the probability is
contained in a single connected window, as opposed to the lower
dimensional scenario where the probability of window occurrence is
distributed over many windows.  We will save further
analysis of this conjecture for a different report (\cite{albersnil}), but hints to why this phenomena is occuring can be found in
\cite{yorkenilpotency}.  


\subsubsection{Relevance}
Decreasing window probability inside the
chaotic region provides direct evidence for conjectures
(\ref{conjecture:windowsize}) and (\ref{conjecture:robustchaos}) along
a one-dimensional interval in parameter space.  We
will, in a more complete manner, attack those conjectures in a different report.  We will use the
decreasing periodic window probability to help verify conjecture
(\ref{conjecture:zerobifvol}) since it provides the context we desire with
the $num-$continuity of the Lyapunov spectrum.  Our argument requires that there exists at least one maximum in the
number of positive Lyapunov exponents with parameter variation.
Further, that maximum must increase monotonically with the dimension
of the system.  The existence of periodic windows causes the following problems:
periodic windows can still yield structural instability - but in a
catastrophic way; periodic windows split up our bifurcation chain sets
which, despite not being terminal to our
arguments, provide many complications with which we do not contend.
However, we do observe a decrease in periodic windows and with
the decrease in the (numerical) existence of periodic windows comes
the decrease in the number of bifurcation chain sets; i.e. $l = |b_n
- a_1|$ is increasing yet will remain finite.

\subsection{Hyperbolocity violation}
We will present two arguments for hyperbolicity violation - or
nearness to hyperbolicity violation of a map at
a particular parameter value, $s$.  The first argument will consider the
fraction of Lyapunov exponents near zero over an ensemble of
networks versus variation in the $s$ parameter.  If there is any hope
of the existence of a chain link set with bifurcation link sets of
decreasing length, our networks (on the $s$ interval in question) must
always have a Lyapunov exponent near zero.  The second argument will come implicitly from $a-$density arguments
presented in section (\ref{sec:adensity}).  To argue for this
conjecture, we only need
the existence of a neutral direction\footnote{By neutral direction we
  mean a zero Lyapunov exponent; we do not wish to imply that there
  will always exist a center manifold corresponding to the zero
  Lyapunov exponent.}, or, more accurately, at least two
bifurcation link sets, which is not beyond reach.

\subsubsection{Qualitative analysis}
A qualitative analysis of hyperbolocity violation comes from combining
the $num-$continuity of the exponents in Fig. (\ref{fig:biglced64}) and
the evidence of exponent zero crossings from Figs.
(\ref{fig:numberoflcepos}) and (\ref{fig:ncversusd}).  If the exponents are
continuous with respect to parameter variation (at least locally) and they start
negative, become positive, and eventually become negative, then they must be zero
(within numerical precision) for at least two points in the parameter
space.  It happens that the bifurcation chain link sets are
LCE decreasing from $i$ to $i+1$, which will provide additional,
helpful, structure.

\subsubsection{Quantitative and numerical analysis}
\begin{figure}
\begin{center}
\epsfig{file=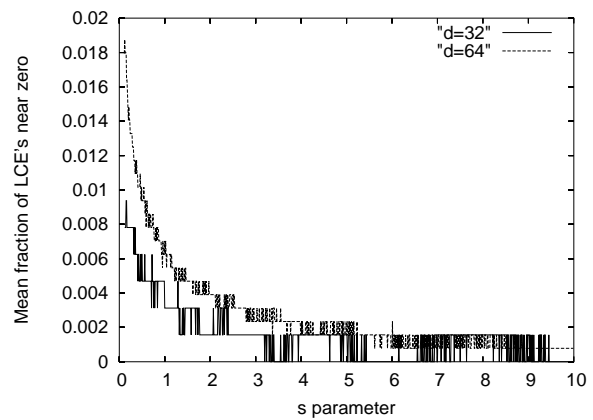, height=8cm, angle=270}
\caption{Mean fraction of LE's near zero ($0 \pm 0.01$) for networks
  with $32$ neurons and $32$ or $64$ dimensions (averaged over $100$ networks).} 
\label{fig:perlcenearzero}
\end{center}
\end{figure}

The first argument, which is more of a necessary but not sufficient
condition for the existence of hyperbolicity violation, consists of searching for the existence of Lyapunov
exponents that are zero within allowed numerical errors.  With
$num-$continuity, this establishes the existence of exponents that are
numerically zero.  For an intuitive feel for what numerically zero
means, consider the oscillations in Fig. (\ref{fig:numberoflcepos}) of the
number of positive exponents versus parameter variation.  It is clear
that as they cross zero there are numerical errors that cause an
apparent oscillation in the exponent; these oscillations are due
largely to numerical fluctuations in the calculations\footnote{It is
  possible that there exist Milnor style attractors for our high-dimensional networks, or at least multiple basins of attraction.  As
this issue seems to not contribute, we will save this discussion for a different report.}.  There is a
certain fuzziness in numerical results that is impossible to remove,
thus questions regarding exponents being exactly zero are ill-formed.
Numerical results of the type presented in this paper need to be
viewed in a framework similar to physical experimental results.  With
this in mind, we need to note the significance of the exponents near zero.  To do
this, we calculate the relative number of Lyapunov exponents numerically
at zero compared to the ones away from zero.  All this information can
be summarized in Fig. (\ref{fig:perlcenearzero}) which addresses the mean fraction of exponents
that are near zero versus parameter variation.

The cut-off for an exponent being near zero is $\pm 0.01$, which is
approximately the expected numerical error in the exponents
for the number of iterations we are using.  There are four important
features to notice about Fig. (\ref{fig:perlcenearzero}): there are no
sharp discontinuities in the curves; there exists an interval in
parameter space such that there is always at least one Lyapunov
exponent in the interval $(-0.01, 0.01)$ and the length of that
parameter interval is increasing with dimension; the curves are concave --- implying that exponents are somehow leaving
the interval $(-0.01, 0.01)$; and there is a higher fraction of
exponents near zero at the same $s$ value for higher dimension.  The
first property is important because holes in the parameter space where
there are no exponents near zero would imply the absence of the
continuous zero crossings we will need to
satisfy conjecture (\ref{conjecture:zerobifvol}).  To satisfy conjecture (\ref{conjecture:hypvol}) we
only need three exponents to be near zero and undergo a zero crossing
for the minimal bifurcation chain subset\footnote{The minimal
  bifurcation chain subset requires at least two adjoining bifurcation
link sets to exist.} to exist.  There are clearly enough exponents on
average for such to exist for at least some interval in parameter
space at $d=32$, e.g. for $(0.1, 0.5)$.  For $d=64$ that interval is much
longer --- $(0.1, 1)$.  Finally, if we want the chain link set to be
more connected and for the distance between elements of the
bifurcation chain subset to decrease, we will need the fraction of
exponents near zero for the fixed interval $(-0.01, 0.01)$ for a given
interval in $s$ to increase with dimension.  This figure does not
imply that there will exist zero-crossings, but it provides the
necessary circumstance for our arguments.

The second argument falls out of the $a-$density and $num-$continuity
arguments.  We know that as the dimension is increased, the variation
of Lyapunov
exponents versus parameter variation decreases until, at dimension $64$, the exponent variation varies continuously within
numerical errors (and thus upon moving through zero, the exponent moves through
zero continuously).  We also know that on the interval in parameter
space $A=[0.1,6]$, the distance between exponent zero crossings
decreases monotonically.  Further, on this subset $A$, there always
exists a positive Lyapunov exponent, thus implying the existence of
bifurcation chain set whose length is at least $5.9$.  Extrapolating these results to their limits
in infinite dimensions, the number of exponent crossings on the
interval $A$ will monotonically increase with dimension.  As
can be seen from Fig. (\ref{fig:adensity}), the exponent zero-crossings
are relatively uniform with the distance between crossings decreasing
with increasing dimension.  Considering Fig. (\ref{fig:poslce4and64}), the exponent zero
crossings are also transverse to the $s$ axis.  Thus the zero crossings on the interval $A$,
which are exactly the
points of non-hyperbolocity we are searching for, are becoming dense.
This is overkill for the verification of the existence of a minimal bifurcation chain set.
This is strong evidence for both conjectures (\ref{conjecture:hypvol})
and (\ref{conjecture:zerobifvol}) .  It is worth noting that
hitting these points of hyperbolocity violation upon parameter
variation is extremely unlikely under any uniform
measure on $R$ as they are a countable collection of points.\footnote{When considering parameter values greater than the
  point where the smallest exponent that becomes positive, becomes
  positive, the zero crossings seem always to be transverse.  For
  smaller parameter values - along the route to chaos, a much more
  complicated scenario ensues.}  Luckily, this does not matter for
either the conjecture at hand or for any of
our other arguments.

\subsubsection{Relevance}
The above argument provides direct numerical evidence of hyperbolocity violation over a
range of the parameter space.  This is strong evidence supporting conjecture
(\ref{conjecture:hypvol}).  It does not yet verify conjecture
(\ref{conjecture:zerobifvol}), but it sets the stage as we have shown that there is a significant range over which hyperbolocity is
violated.  The former statement speaks to conjecture
(\ref{conjecture:nongenercityss}) also; a full explanation of conjecture
(\ref{conjecture:nongenercityss}) requires further analysis, which is
the subject of a discussion in the final remarks.

\subsection{Hyperbolocity violation versus parameter variation}

We are finally in a position to consider the final arguments for
conjecture (\ref{conjecture:zerobifvol}).  To complete this analysis, we
will need the following pieces of information:
\begin{itemize}
\item[i.] we need the maximum number of positive exponents to go to infinity

\item[ii.] we need a region of parameter space for which $a-$density
  of Lyapunov exponent zero crossings exists; i.e. we need an
  arbitrarily large number of adjoining bifurcation link sets (such
  that the cardinality of the bifurcation chain set becomes
  arbitrarily high) such that for each  $V_i$, the length of $V_i$, $l = |b_i - a_i|$, approaches zero. 

\item[iii.] we need $num-$continuity of exponents to increase as the dimension increases

\item[iv.] a major simplification can be provided with the existence of one global maximum in the number of positive
exponents and entropy, and along any portion of parameter space where $s$
is greater than the $s$ at the maximum number of positive exponents, the
maximum and minimum number of exponents occur on the graph at the end points of the parameter range
(within numerical accuracy)

\end{itemize}

The $a-$density, $num-$continuity and the arbitrary numbers of positive
exponent arguments we need have, for the most
part, been provided in previous sections.  In this section we will
simply apply the $a-$density and $num-$continuity results in a manner
that suits our needs.  The evidence for the existence of a single maximum in the number of
positive exponents, a mere convenience for our presentation, is evident
from section (\ref{sec:arbnumberofposexponents}).  We will simply rely on all our previous figures and the
empirical observation that as the dimension is
increased above $d=32$, for networks that have the typical
$num-$continuity (which includes all networks observed for $d\geq
64$), there exists a single, global maximum in the number of
positive exponents versus parameter variation.

\subsubsection{Qualitative analysis}
The qualitative picture we are using for intution is that of Fig. (\ref{fig:poslce4and64}).
This figure displays all the information we wish to quantify for many
networks; as the dimension is increased, there is a region of
parameter space where the parameter variation needed to achieve a
topologically different (by topologically different, we mean a
different number of global stable and unstable manifolds) attractor decreases to zero.  Based on Fig.
(\ref{fig:poslce4and64}) (and hundreds of similar plots), we claim
that qualitatively this parameter range exists for at least $0.5
\leq s \leq 6$.

\subsubsection{Quantitative and numerical analysis}
Let us now complete our arguments for conjecture
(\ref{conjecture:zerobifvol}).  For this we need a subset of the
parameter space, $B \subset R^1$, such that some variation of $s \in B$
will lead to a topological change in the map $f$ in the form of a
change in the number of global stable and unstable manifolds.
Specifically, we need $B = \cup V_i = V$, where $V_i$ and $V_{i+1}$ share
a limit point and are disjoint.  Further, we need the variation in $s$ needed for the topological change
to decrease monotonically with dimension on $V$.  More precisely, on
the bifurcation chain set, $U$, the distance between elements must decrease
monotonically with increasing dimension.  We will argue in
three steps: first, we will argue that, for each $f$ with a sufficiently
high number of dimensions, there will exist an arbitrarily large
number of exponent zero crossings (equivalent to an arbitrarily large number of positive exponents); next we will argue
that the zero crossings are relatively smooth; and finally, we will argue
that the zero crossings form an $a$-dense set on $V$ --- or on the
bifurcation chain set, $l = |b_i - a_i| \rightarrow 0$ as $d
\rightarrow \infty$.  This provides
strong evidence supporting conjecture (\ref{conjecture:zerobifvol}).  

Assume a sufficiently large number of dimensions, verification of
conjecture (\ref{conjecture:hypvol}) gives us the existence of the
bifurcation chain set and the existence of the adjoining bifurcation
link sets.  The existance of an arbitary number of positive Lyapunov,
and thus an arbitrarily large number of zero crossings follows from section
(\ref{sec:arbnumberofposexponents}).  That the bifurcation chain set has
an arbitrarily large number of elements, $\# U
\rightarrow \infty$ is established by conjecture
(\ref{conjecture:windowsize}), because, without periodic windows,
every bifurcation link set will share a limit point with another
bifurcation link set.  From section
(\ref{sec:ncontinuity}), the $num-$continuity of the exponents persists
for a sufficiently large number of dimensions, thus the Lyapunov
exponents will cross through zero.  Finally, section
(\ref{sec:adensity}) tells us that the Lyapunov exponent zero
crossings are $a-$dense, thus, for all $c_i \in U$, $|c_i - c_{i+1}|
\rightarrow 0$, where $c_i$ and $c_{i+1}$ are sequential elements of
$U$.

Specifically for our work, we can identify $U$ such that $U \subset
[0.5, 6]$.  We could easily extend the upper bound to much greater than
$6$ for large dimensions ($d\geq 128$).  How high the upper bound can
be extended will be a discussion in further work.  

Finally, it is useful to note that the bifurcation link sets are LCE
decreasing with increasing $s$.  This is not necessary to our
arguments, but it is a nice added structure that aids our intuition.
The LCE decreasing property exists due to the existence of the single, global maximum in
the maximum number of positive Lyapunov exponents followed by an apparent exponential fall off in the
number of positive Lyapunov exponents.

\subsubsection{Relevance}
The above arguments provide direct evidence of conjectures
(\ref{conjecture:zerobifvol}) and (\ref{conjecture:nongenercityss})
for a one-dimensional curve (specifically an interval) in parameter
space for our networks.  This evidence also gives a hint with respect
to the robustness of chaos in high-dimensional networks with
perturbations on higher-dimensional surfaces in parameter space.
Finally, despite the seemingly inevitable topological change upon
minor parameter variation, the topological change is quite benign.

\section{Fitting everything together}
\label{sec:summary}

Having finished with our specific analysis, it is now time to put our
work in the context of other work, both of a more mathematical and a more
practical and experimental nature.  In this spirit, we will provide, first, a brief summary of our arguments followed by a discussion of how our
results fit together with various theoretical results from dynamical
systems and turbulence.  

\subsection{Summary of arguments}
We will give brief summaries of our
results, both in the interest of clarity and to relate our
results and methods to others.  

\subsubsection{Periodic window probability decreasing: conjecture \ref{conjecture:windowsize}}
The conjecture that the probability of periodic window existence for a
given $s$ value along an interval in parameter space decreases with
increasing dimension upon the smallest positive Lyapunov exponent
becoming positive, is initially clear from
considering the Lyapunov spectra of neural networks versus parameter
variation for networks of increasing size (Figs.
(\ref{fig:biglced4}) and (\ref{fig:biglced64})).  We show that as
the dimension is increased, the observed probability of periodic
windows decreases inversely with increase in dimension.  The motivation for arguing in
this way is simple; this analysis is independent from the
$num-$continuity analysis, and the results from the analysis of
$num-$continuity and periodic window probability decrease reinforce each
other.  The mechanism that this conjecture provides us with is the
lengthening of the bifurcation chain set.

Further investigations of this particular phenomena will follow in a
later report.  For other related results see \cite{yorkenilpotency}, \cite{yorkerobustchaos},
\cite{nnrobustchaos}, and \cite{albersnil}. 

\subsubsection{Hyperbolicity violation: conjecture \ref{conjecture:hypvol}}

The intuition for this conjecture arises from observing that for our high-dimensional systems,
there exists at least one Lyapunov exponent that starts negative, becomes positive,
then goes negative again; thus if it behaves numerically continuously, it
must pass through zero for some parameter value $s$.

To verify this conjecture, we presented two different arguments.  This
first argument was a necessary but not sufficient condition for
hyperbolicity violations.  We show that over a sizeable interval in
parameter space, there exists a Lyapunov exponent very near zero, and
the fraction of the total number of Lyapunov exponents that are
near zero increases over a larger interval of parameter space as the
dimension is increased.  The second argument was based on the $a-$density of exponent zero
crossings, the $num-$continuity of the exponents as the dimension
increased, and the increasing number of positive exponents with
dimension.  Both arguments together help imply an interval of
parameter space such that on that interval, the number of parameter
values such that hyperbolicity is violated is increasing.

\subsubsection{Existence of Codimension-$\epsilon$ bifurcation set: conjecture \ref{conjecture:zerobifvol}}
Conjecture (\ref{conjecture:zerobifvol}) is the next step in relating
our results with the results of structural stability theory.  Given
the results supporting conjecture (\ref{conjecture:hypvol}), conjecture (\ref{conjecture:zerobifvol}) only needs a few added bits of
evidence for its vindication.  

The intuition for this argument follows from observing that the
peak in the number of positive Lyapunov exponents tends toward a spike
of increasing height and decreasing width as the dimension is increased.  This, with some sort of
continuity of exponents, argues for a decrease in distance between
exponent zero crossings.

A summary for the arguments regarding conjecture
(\ref{conjecture:zerobifvol}) is as follows.  With increasing dimension we have: increased $num-$continuity of Lyapunov exponents;
increasing number of positive Lyapunov exponents; and $a-$density of
Lyapunov exponent zero crossings (thus all the exponents are not
clustered on top of each other).  Thus, on a finite set in parameter
space, we have an arbitrary number of exponents that move numerically
smoothly from negative values, to positive values, and back to negative
values.  Further, these exponents are relatively evenly spaced.  Thus,
the set in parameter space for which hyperbolicity is
violated is increasingly dense; and
with an arbitrarily number of violations available, the
perturbation of the parameter required to force a topological change
(a change in the number of positive exponents) becomes small.

\subsubsection{Non-genericity of structural stability: conjecture \ref{conjecture:nongenercityss}}
\label{sec:genssconclusion}
As previously mentioned, it could appear that our results are contrary to  Robbin \cite{robbinss}, Robinson
\cite{robinsonss1}, and Ma\~n\'e \cite{maness}.  We will discuss
specifically how our results fit with theirs in section
(\ref{sec:usandss}).  In the current discussion, we wish to properly interpret
our results in a numerical context.

We claim to have found a subset of parameter space that, in the limit
of infinite dimensions, has dense hyperbolicity violation.  This could
be interpreted to imply that we have located a set for which
strict hyperbolicity does not imply structural stability, because the $C^1$
changes in the parameter give rise to topologically different behaviors.
The key issue to realize is that in numerical simulations, there do
not exist infinitesimals or infinite-dimensional limits\footnote{Note that in previous sections we do use words
  like, ``in the infinite-dimensional limit.''  It is for reasons
  such as these (and many others) that we are only putting forth
  conjectures and not theorems; this distinction is not trivial.}.  Rather, we can speak to how behaviors arise, and how
limits behave along the path to the ideal.  We have found a subset of
parameter space that we believe can approximate (with unlimited
computing) arbitrarily closely a set for which hyperbolicity will not
imply structural stability.  Thus, an experimentalist or a numerical
physicist might see behavior that looks like it violates the results
of Robbin \cite{robbinss}, Robinson
\cite{robinsonss1}, and Ma\~n\'e \cite{maness}; yet it will not strictly
be violating those theorems.  The key point of this  conjecture is
that we can observe apparent violation of  the structural stability
conjecture, but the violation (on a Lebesgue measure zero set) occurs as smooth, not
catastrophic, topological change.  (In section (\ref{sec:usandss}) we
will further discuss our results as they relate to those of Robbin \cite{robbinss}, Robinson
\cite{robinsonss1}, and Ma\~n\'e \cite{maness}.)

\subsubsection{Robust chaos: conjecture \ref{conjecture:robustchaos}}
That chaos is a robust behavior for bounded high-dimensional dynamical
systems is not particularly surprising, especially in light of Fig.
(\ref{fig:lcevsiteration16and64}), information presented in sections
(\ref{sec:arbnumberofposexponents}) and (\ref{sec:decreasingwindowsize}),
and previous work \cite{albers1}.  Beyond casual observation, we will
not comment because it is the topic of a work in progress
\cite{albersnil}.  It is important to note that we do not observe
sinks or periodic windows in the chaotic region of parameter space for
a sufficiently high dimension.  This particular
characteristic is, however, somewhat heartening if one is to compare our
results with many high-dimensional chaotic and turbulent natural
systems as these systems are constantly being perturbed, yet their
behavior is relatively robust.  Readers interested in arguments along
the lines of conjecture (\ref{conjecture:robustchaos}) are directed to
\cite{yorkenilpotency}, \cite{yorkerobustchaos}, \cite{nnrobustchaos},
\cite{albersnil}, or \cite{unimodalrobust} for further information.

\subsection{Fitting our results in with the space of $C^r$ function:
  how our network selection method affects our view of function space}
Performing a numerical experiment
induces a measure upon whatever is being experimented upon.  We now discuss some of the characteristics of our imposed
measure and how they might affect our results.  Recall, often in
mathematics, it is desirable to prove that various results are
invariant to the measure imposed upon the space; in our case this
would be extremely difficult if not impossible, thus we will resort to
the aforementioned, standard experimental style.

A measure, in a very general sense, provides a method of measuring the
volume a set occupies in its ambient space (for a formal treatment,
see \cite{whezyg}).  Usually that method
provides a specific mechanism of measuring lengths of a covering
interval.  Then, the entire space is covered with the aforementioned
intervals, and their collective volume is summed.  One of
the key issues is how the intervals are weighted.  For instance,
considering the real line with the standard Gaussian measure imposed upon it;
the interval $[-1, 1]$ contains the majority of the volume of the
entire interval $[-\infty, \infty]$.  Our method of weighting networks
selects fully connected networks with random Gaussian weights.
Thus, in limit of high dimension and high number of neurons, very
weakly connected networks will be rare, as the Gaussian statistics of
the weights will be dominant.  Likewise, fully connected networks
where all the weights have the same strength (up to an order of
magnitude) will also be uncommon.  One can argue whether our measure
realistically represents the function space of
nature, but those arguments are ill-formed because they cannot be answered
without either specific information about the natural system with
which our
framework is being compared, or the existence of some type of
invariant measure.  Nevertheless, our framework does cover the
entire space of neural networks noted in section
(\ref{sec:dsfunctionspace}), although all
sets do not have equal likelihood of being selected, and thus our
results must be interpreted with this in mind.

A second key issue regards how the ambient space is split into
intervals; or in a numerical sense, how the grain of the space is
constructed.  We will again introduce a simpler case for purposes of
illustration, followed by a justification of why the simpler case and
our network framework are essentially equivalent.  Begin with $R^n$
and select each coordinate ($v_i)$ in the vector $\bf{v} = \{ v_1, v_2, \cdots,
v_n \} \in R^n$ from a normal, $\mathit{i.i.d.}$ distribution with mean zero, variance
one.  Next, suppose that we are attempting to see every number and
every number combination.  This will be partially achieved by the
random number selection process mentioned above, and it is further
explored by sweeping the variance, i.e. selecting a scalar $s \in R$,
$0 \leq s$, and sweeping $s$ over the positive real line, $s \textbf{v} $.
This establishes two meshes, one for the individual vectors which is
controlled by how finely the $s$ parameter is varied, and another mesh
that controls how the initial coordinates are selected.  These two
combined meshes determine the set of combinations of coordinates that will be
observed.  If one considers how this affects vector
selection in, say, $R^3$ for simplicity, both in the initial vector selection and in the
vector sweeping, it is clear how $R^3$ will be carved out.

The point of the above paragraph is simply this: we are associating
how we carve up our neural network function space with how we carve up
the neural network weight space.  It should be clear that this is
comparing apples to apples.  In the above paragraph, to understand how
our neural network selection process works, simply associate $\textbf{v}$
with the vectors in the $w$ matrix and the scaling parameter $s$ with
$s$.  This keeps the view of our function space largely in standard
Euclidean space.  Of course there is the last remaining issue of the
amplitude terms, the $\beta$'s.  Apply the same type of analysis
to the $\beta$'s as we did for the $w$'s in the above paragraph.  Of
course initially it would seem that the scaling parameter is missing,
but note that multiplying the $\beta$'s by $s$, in our networks, is
essentially equivalent to multiplying the $w$'s by $s$.  To understand
this, consider the one-dimensional network, with one neuron:
\begin{equation}
x_{t+1} = \beta_0 + \beta_1 \tanh(sw_0 + sw_1(\beta_0 + \beta_1
\tanh(sw_0 + sw_1x_{t-1})))
\end{equation} 
It is clear from this that inserting $s$ inside $\tanh$ will sweep the $\beta$'s, but inserting $s$ outside the squashing
function will miss sweeping the $w_0$ bias term.\footnote{Recalling
  section (\ref{sec:ournnconstruction}), we actually do a little more
  with the $\beta$'s than we mention here; the previous argument is
  simply meant to give a (mathematically imprecise) picture of how our
  experiment carves out the space of neural networks.}

From this is should be clear that our framework will capture the
entire space of neural networks we are employing.  Yet, it should also
be clear that we will not select each network with equal
probability.  Weakly connected networks will not be particularly
common in our study, especially as the number of dimensions and
neurons increase, because the statistics of our weights will more closely
resemble their theoretical distributions.  It is also worth noting
that a full connection between network structure and dynamics, in a
sensible way, is yet out of reach (as opposed to, say, for spherical harmonics).  Nevertheless, we claim that
our framework gives a complete picture of the space of $C^r$ maps of
compact sets to compact sets with the Sobolev metric from the
perspective of a particular network selection method.

\subsection{Our results related to other results in dynamical systems}
As promised throughout, we will now connect our results with various
theorems and conjectures in the field of dynamical systems.  This will
hopefully help put our work in context and increase it's
understandability.  We will address how our work fits in
with the stability conjecture of Smale and Palis \cite{globalanalysisss}.   First we will
discuss our results and the structural
stability theories of Robbin \cite{robbinss} and Ma\~n\'e \cite{maness} which state that structurally
stable systems must be hyperbolic.  We will follow this by relating our
studies to the work in partial hyperbolicity and stable ergodicity -- the
reaction to difficulties in showing that hyperbolic systems are
structurally stable.  We will conclude this portion of the summary by
discussing how our work relates to one of the conjectures from a paper by Palis \cite{palisconjecture}.


\subsubsection{Structural stability theory
 and conjecture \ref{conjecture:nongenercityss}}
\label{sec:usandss}
It is now time to address the apparent conflict between our
observations and the structural stability
theorems of Robbin \cite{robbinss}, Robinson \cite{robinsonss1}, and
Ma\~n\'e \cite{maness}.  We would like to begin by noting that we do not
doubt the validity or correctness of any of the aforementioned
results.  In fact, any attempt to use our techniques and results to
provide a counter example to the theorems of Robbin, Robinson, or Ma\~n\'e
involves a misunderstanding of what our methods are able
to do and indeed intend to imply.  

In conjecture (\ref{conjecture:nongenercityss}) we claim, in an intuitive
sense, that along a one-dimensional curve in parameter space, our dynamical systems are
hyperbolic with measure one, with respect to Lebesgue measure.  Yet,
we can still find subsets that are measure zero, yet $a-$dense, for which our dynamical systems are
partially hyperbolic rather than hyperbolic.  The motivation for the
above statement roughly derives from thinking of a turbulent fluid.  In this circumstance, the number
of unstable manifolds can be countably infinite, and upon varying,
say, the viscosity, from very low to very high, one would have a countable
number of exponents becoming positive over a finite length of
parameter space.  Yet, all the limits of this sort and all the intuitive
ideas with respect to what will happen in the infinite-dimensional
limit, are just that, ideas.  There are limits to what we can compute;
there do not exist infinite-dimensional limits in numerical computing;
there do not exist infinitesimals in numerical computing; and aside
from the existence of convergence theorems, we are left unable to draw
conclusions beyond what our data says.  Thus, our results do not
provide any sort of counter example to the stability conjecture.
Rather, a key point of our results is that we do observe, in a
realistic numerical setting, structural instability upon small
parameter variation.  It is useful to think instead of structural
stability as an open condition on our parameter space whose endpoints
correspond to the points  of structural instability - the  points of
bifurcations in turbulence.  These disjoint open sets are precisely
the bifurcation link subsets, $V_i$ for which the map $f$ is
structurally stable.  As the dimension is increased,
the length of the $V_i$'s decreases dramatically, and
may fall below numerical or experimental resolution.  Thus, the
numerical or experimental scientist might observe, upon parameter
variation, systems that should according to the work of Robbin,
Robinson and Ma\~n\'e, be structurally stable, to undergo topological
variation in the form of a variation in the number of positive
Lyapunov exponents; i.e. the scientist might observe structural instability.
This is the very practical difference between numerical computing and
the world of strict mathematics.  (Recall we were going
to attempt to connect structural stability theory closer to reality,
the former statement is as far as we will go in this  report.)  The
good news is that even though observed structural stability might be
lost, it is lost in a very meek manner - the topological changes are
very slight, just as seems to be observed in many turbulent experimental systems.  Further, partial hyperbolicity is not lost, and
the dynamically stable characteristics of stable ergodicity seem to be
preserved, although we obviously can't make a strict mathematical statement.  

Thus, rather than claiming our results are contrary to those of  Robbin \cite{robbinss}, Robinson \cite{robinsonss1}, and
Ma\~n\'e \cite{maness}, we note that our results speak both to what might
be seen of those theorems in high-dimensional dynamical systems and
how their results are approached upon increasing the dimension of a
dynamical system.  

It is worth noting that, given a typical $64$-dimensional network, if
we fixed $s$ at such a point that there was an exponent zero crossing,
we believe (based upon preliminary results) that there will exist many
perturbations of other parameters that leave the exponent zero
crossing unaffected.  However, it is believed at this time that these
perturbations are of very small measure (with respect to Lebesque measure), and of a small codimension
set, in parameter space, i.e. we believe we can find perturbations
that will leave the seemingly transversal intersection of an exponent
with $0$ at a particular $s$ value unchanged, yet these parameter
changes must be small.

\subsubsection{Partial hyperbolicity}

In this study we are particularly concerned with the
interplay, along a parameterized curve, of how often partial
hyperbolicity is encountered versus strict hyperbolicity.  It should be
noted that if a dynamical system is hyperbolic, it is partially
hyperbolic.  All of the neural networks we considered were at least
partially hyperbolic; we found no exceptions.  Many of the important
questions regarding partially hyperbolic dynamical systems lies in
showing the conditions under which such systems are stably ergodic.
We will now discuss this in relation to our results and methods.

Pugh and Shub \cite{parthyppughshub2} put forth the following conjecture regarding partial
hyperbolicity and stable ergodicity:
\begin{conjecture}{ \bf (Pugh and Shub \cite{parthyppughshub2} Conjecture
    $3$)}
\label{conjecture:pughshubstabilityconjecture}
Let $f \in \mathit{Diff}^2_{\mu}(M)$ where $M$ is compact.  If $f$ is
partially hyperbolic and essentially accessible, then $f$ is ergodic.
\end{conjecture}
In that same paper they also proved the strongest result that had
been shown to date regarding their conjecture:
\begin{theorem}{ \bf (Pugh-Shub theorem (theorem $A$ \cite{parthyppughshub2}))}
\label{theorem:pughshub}
If $f \in \mathit{Diff}_{\mu}^2(M)$ is a center bunched, partially
hyperbolic, dynamically coherent diffeomorphism with the essential
accessibility property, then $f$ is ergodic.
\end{theorem}
A diffeomorphism is partially hyperbolic if it satisfies the
conditions of definition (\ref{definition:partialhyperbolicity}).  Ergodic behavior implies that, upon breaking the attractor into
measurable sets, $A_i$, for $f$ applied to each measurable set for enough
time, $f^n(A_i)$ will intersect every other measurable set $A_j$.  This implies a weak
sense of recurrence; for instance, quasi-periodic orbits, chaotic orbits,
and some random processes are at least colloquially ergodic.  More formally, a dynamical
system is ergodic if and only if almost every point of each set
visits every set with positive measure.  The accessibility property
simply formalizes a notion of one point being able to reach another
point.  Given a partially hyperbolic dynamical system, $f: X
\rightarrow X$ such that there is a splitting on the tangent bundle
$TM = E^u \oplus E^c \oplus E^s$, and $x, y \in X$, $y$ is accessible
from $x$ if there is a $C^1$ path from $x$ to $y$ whose tangent
vector lies in $E^u \cap E^s$ and vanishes finitely many times.  The
diffeomorphism $f$ is center bunched if the spectra of $Tf$ (as
defined in section (\ref{sec:lceandhyperbolicity})) corresponding to the
stable ($T^sf$), unstable ($T^uf$), and ($T^cf$) central directions
lie in thin, well separated annuli (see \cite{parthyppughshub}, page
131 for more detail, the
radii of the annuli is technical and is determined by the Holder
continuity of the diffeomorphism.)  Lastly, let us note that a dynamical system is called stably ergodic
if, given $f \in \mathit{Diff}^2_{\mu}(M)$ (again $M$ compact), there is a
neighborhood, $f \in Y \subset \mathit{Diff}^2_{\mu}(M)$ such that every $g \in
Y$ is ergodic with respect to $\mu$.  We will refrain from divulging
an explanation of dynamical coherence; it is a very crucial
characteristic for the proof of theorem (\ref{theorem:pughshub}), but we
will have little to say in its regard.

An actual numerical verification of ergodicity can be somewhat
difficult as the modeler would have to watch each point and verify
that eventually the trajectory returned very close to every other
point on the orbit (i.e. it satisfies the Birkoff hypothesis).
Doing this for a few points is, of course, possible, but doing it for a
high-dimensional attractor for any sizable number of points can be
extremely time consuming.  Checking the accessibility criterion seems
to pose similar problems - in fact it is hoped that accessibility is the
sufficient recurrence conditions for ergodic behavior - thus is should
be no surprise that accessibility would be difficult to check
numerically (it has been shown to be $C^1$ dense \cite{essaccc1dense}).  In the reality of computing, there is a far
more practical way of checking for ergodic behavior, motivated by a
more practical problem in numerical computing, transients.  For a
mathematician, ergodic tools can be applied whenever the system can be
shown to be ergodic.  In numerical work, proving that the necessary
conditions for the use of ergodic measures is often intractable.
Besides, for numerical applications, proving long-term
behavior is often not
good enough since the use of an ergodic diagnostic, for the relaxation from the transients to the ergodic
state can, at times, be prohibitively slow, and sometimes difficult to
detect.  There are even times when the numerical errors in the
calculations effectively reset the transients.  The practical solution
to this is to apply the  ergodic measures and, along with the time-series data, watch the transients disappear.  We did this specifically
in section (\ref{sec:numericalerror}) to justify our use of ergodic
measures.  If the errors in the ergodic measures along with the
transients of the attractors decrease with time, then we call the
system ergodic and feel justified in using ergodic measures, such as
Lyapunov exponents.  

Considering Figs. (\ref{fig:lcevsiteration16and64}),
(\ref{fig:closeup64}), and (\ref{fig:meanlceerror16and64}),
it seems clear that our networks are ergodic since the ergodic measures
converge.  Further, upon considering Figs. (\ref{fig:biglced4}),
(\ref{fig:biglced64}), and (\ref{fig:poslce4and64}), when a one-dimensional
parameter is varied, ergodic behavior is preserved.  Of course,
showing that one has explored all the variations inside the
neighborhood $(f \in ) Y \subset \mathit{Diff}^2_{\mu}(M)$ is impossible: thus
claiming that we have, in a mathematically rigorous way, observed stable ergodicity as the predominant characteristic would be
premature.  Further, we can say little about the accessibility
property.  What we can say is that we have never observed a dynamical
system, within our construction, that is not on a compact set, is not
partially hyperbolic, and is not stably ergodic.  Thus, our results
provide evidence that the conjecture of Shub and Pugh is on track.
For more information with respect to the mathematics discussed above,
see \cite{parthyppughshub}, \cite{reviewstablergodicity}, or \cite{partialhyplese}.
 
Comparing conjecture (\ref{conjecture:pughshubstabilityconjecture}) to
theorem (\ref{theorem:pughshub}), the required extra hypotheses for the
proof of the theorem are dynamical coherence and center bunching of
the spectrum of $Tf$.  Pugh and Shub, and others have been attempting
to eliminate these extra hypothesis.  Our results speak little to the
issue of dynamic coherence, but our results can speak to the issue of
center bunching.  Considering Fig. (\ref{fig:biglced64}) at any value
of the $s$ parameter, there is no evidence of center bunching, or any
real bunching of Lyapunov exponents at all.  In fact, if there were
center bunching, our $a-$density of exponent zero crossing argument
would be in serious trouble.  Thus, we claim that we have strong
evidence for the removal of the center bunching requirement for stable
ergodicity.  And, since we are claiming that our
dynamical systems are seemingly ergodic, if center bunching were
required of stable ergodicity, we claim that stable ergodicity would
be too strict of a distinguishing characteristic for dynamic
stability.\footnote{It is worth noting that in section $31$ of Landau
  and Lifshitz's fluid mechanics book \cite{llfluid}, they give
  physical reasons why one might expect a center bunching type of
  spectrum, or at least a finite number of exponents near zero, in
  turbulent fluids.}

\subsubsection{Our results and Palis's conjectures}
Palis \cite{palisconjecture} stated many stability conjectures
based upon the last thirty years of developments in dynamical
systems.  We wish to address one of his conjectures:
\begin{conjecture}{ \bf (Palis \cite{palisconjecture} conjecture $II$)}
In any dimension, the diffeomorphisms exhibiting either a homoclinic
tangency or a (finite) cycle of hyperbolic periodic orbits with
different stable dimensions (heterodimensiopnal cycle) are $C^r$ dense
in the complement of the closure of the hyperbolic ones. 
\end{conjecture}
Let us decompress this for a moment, and then discuss how our results
fit with it.  Begin by defining the space of $d$-dimensional $C^r$ diffeomorphisms as $X$.  Next, break that space up
as follows: $A = \{ x \in X | x $ exhibits a homoclinic tangency or a
finite cycle of hyperbolic periodic orbits with different stable
dimensions $\}$ and $B = \{ x \in X | x \text{ is hyperbolic } \}$.
Thus $B$ is the set of hyperbolic, aperiodic diffeomorphisms, and $A$
is the set of periodic orbits or partially hyperbolic orbits.  The
conjecture states that $A$ is dense in the complement of the closure
of $B$; thus $A$ can be dense in $B$.  With respect to our results,
the partially hyperbolic diffeomorphisms (diffeomorphisms with
homoclinic tangencies) can be dense within the set of hyperbolic
diffeomorphisms.  Our conjectures claim to find a subset of our one-dimensional parameter space such that partially hyperbolic
diffeomorphisms will, in the limit of high dimensions, be dense.  In
other words, our work not only agrees with Palis's
conjecture $II$ (and subsequently his conjecture $III$), but our work provides
evidence confirming Palis's conjectures.  Of course, we do not
claim to provide mathematical proofs, but rather strong numerical evidence
supporting Palis's ideas.

\subsection{Final remarks}

Finally, let us briefly summarize:
\begin{sor}[(Summary)]
Assuming our particular conditions and our particular space of $C^r$
dynamical systems as per section \ref{sec:nnfunctionspace}, there
exists a collection of bifurcation link subsets ($V$) such that, in the limit of countably infinite
dimensions, we have numerical evidence for the following:

Conjecture \ref{conjecture:hypvol}: on the above mentioned set $V$, strict
  hyperbolicity will be violated $a-densely$.

Conjecture \ref{conjecture:zerobifvol}: on the above mentioned set $V$, the number of
  stable and/or unstable manifolds will change under parameter variation
  below numerical precision.

Conjecture \ref{conjecture:windowsize}: on the above mentioned set
  $V$, the probability of the existence of a periodic window for a
  give $s$ on a specific parameter interval decreases inversely with dimension.

Conjecture \ref{conjecture:nongenercityss}: on the above mentioned set $V$, hyperbolic dynamical systems are not structurally stable
  within numerical precision with measure one with respect to Lebesque measure in parameter
  space. 

\end{sor}
In a measure-theoretic sense hyperbolic systems
occupy all the space, but the partially hyperbolic dynamical systems
(with non-empty center manifolds) can be
$a-$dense on $V$.  Intuitively, if there are countable
dimensions - thus countable Lyapunov exponents, then one of two things
can happen upon parameter variation:
\begin{itemize}
\item[i.]  there would have to be a persistent homoclinic tangency- or some other
sort of non-transversal intersection between stable and unstable
manifolds that was persistent to parameter changes;

\item[ii.]  there can be, at most, countable parameter points such that there
are non-transversal intersections between stable and unstable
manifolds. 
\end{itemize}

We also see that for our networks, each exponent in the spectrum
converges to a unique (within numerical resolution) value.  This both
confirms the usefulness and validity of our techniques, and provides
strong evidence for the prevalence of ergodic behavior.  Further, upon
parameter variation, the ergodic behavior is seemingly preserved; thus we also
have strong evidence of a prevalence of stable ergodic behavior.

\section{Acknowledgments}
The authors would like to thank J. R. Albers, D. Feldman, C. McTague,
J. Supanich, and J. Thompson for a careful and thoughtful
reading of this manuscript.  Comments provided by the K. Burns were
indispensable and aided in the correctness and readability of this manuscript.  D. J. Albers would like to
thank R. A. Bayliss, K. Burns, W. D. Dechert, D. Feldman, C. McTague, J. Robbin, C. R. Shalizi,
and J. Supanich for many helpful discussions and advice.  D. J. Albers
would like to give special thanks to J. P. Crutchfield for much
guidance, many fruitful discussions, and support throughout this
project --- his support and insight aided significantly to this report.  The computing for the project was done
on the Beowulf cluster at the Santa Fe Institute and was partially
supported at the Santa Fe Institute under the Networks, Dynamics
Program funded by the Intel Corporation under the Computation,
Dynamics, and Inference Program via SFI's core grant from the National
Science and MacArthur Foundations.  Direct support for D. J. Albers
was provided by NSF grants DMR-9820816 and PHY-9910217 and DARPA
Agreement F30602-00-2-0583.



\bibliography{dstexts,partialhyperbolicity,lyapunovexponents,neuralnetworks,nilpotency,topology,analysis,structuralstability,computation,me,physics,probability}


\end{document}